\documentclass[%
aip,
 amsmath,amssymb,
 reprint,%
tightenlines,
flushbottom,
citeautoscript,
]{revtex4-1}

\usepackage{graphicx}
\usepackage{dcolumn}
\usepackage{bm}

\usepackage{booktabs}
\usepackage{ wasysym }
\usepackage{siunitx}
\usepackage[utf8]{inputenc}
\usepackage[T1]{fontenc}
\usepackage{mathptmx}
\usepackage{etoolbox}
\usepackage{hyperref}
\hypersetup{
    colorlinks = true,
    linkcolor = blue,
    citecolor = blue,
    urlcolor=blue,
}

\usepackage{tabularx}

\makeatletter
\def\@email#1#2{%
 \endgroup
 \patchcmd{\titleblock@produce}
  {\frontmatter@RRAPformat}
  {\frontmatter@RRAPformat{\produce@RRAP{*#1\href{mailto:#2}{#2}}}\frontmatter@RRAPformat}
  {}{}
}%
\makeatother

\usepackage{setspace}
  \usepackage{parskip}
\usepackage[compact]{titlesec}         
\titlespacing{\section}{4pt}{4pt}{4pt} 
\AtBeginDocument{
  \setlength\abovedisplayskip{4pt}
  \setlength\belowdisplayskip{4pt}
  }
  
  \usepackage[]{caption, subfig}
\usepackage{natmove}
\begin{document}

\preprint{AIP/123-QED}

\title{Radiatively Cooled Magnetic Reconnection Experiments Driven by Pulsed Power}
\author{R. Datta}
\affiliation{Plasma Science and Fusion Center, Massachusetts Institute of Technology, MA 02139, Cambridge, USA\looseness=-10000 
}%
\author{K. Chandler}
\affiliation{ 
Sandia National Laboratories, Albuquerque, NM 87123-1106, USA
}%
\author{C.E. Myers}
\altaffiliation[Current address: ]{Commonwealth Fusion Systems, Devens, MA 01434, USA}
\affiliation{ 
Sandia National Laboratories, Albuquerque, NM 87123-1106, USA
}

\author{J. P. Chittenden}
\affiliation{ 
Blackett Laboratory, Imperial College London, London SW7 2BW, UK\looseness=-10000 
}%
\author{A. J. Crilly}
\affiliation{ 
Blackett Laboratory, Imperial College London, London SW7 2BW, UK\looseness=-10000 
}%

\author{C. Aragon}
\affiliation{ 
Sandia National Laboratories, Albuquerque, NM 87123-1106, USA
}%

\author{D. J. Ampleford}
\affiliation{ 
Sandia National Laboratories, Albuquerque, NM 87123-1106, USA
}%

\author{J. T. Banasek}
\affiliation{ 
Sandia National Laboratories, Albuquerque, NM 87123-1106, USA
}%

\author{A. Edens}
\affiliation{ 
Sandia National Laboratories, Albuquerque, NM 87123-1106, USA
}%
\author{W. R. Fox}
\affiliation{ 
Princeton Plasma Physics Laboratory, Princeton, NJ 08543, USA 
}%
\author{S. B. Hansen}
\affiliation{ 
Sandia National Laboratories, Albuquerque, NM 87123-1106, USA
}%
\author{E. C. Harding}
\affiliation{ 
Sandia National Laboratories, Albuquerque, NM 87123-1106, USA
}%
\author{C. A. Jennings}
\affiliation{ 
Sandia National Laboratories, Albuquerque, NM 87123-1106, USA
}%
\author{H. Ji}
\affiliation{ 
Princeton Plasma Physics Laboratory, Princeton, NJ 08543, USA 
}%
\author{C. C. Kuranz}
\affiliation{Department of Nuclear Engineering and Radiological Sciences, University of Michigan, Ann Arbor, MI 48109, USA\looseness=-1 
}%
\author{S. V. Lebedev}
\affiliation{ 
Blackett Laboratory, Imperial College London, London SW7 2BW, UK\looseness=-10000 
}%

\author{Q. Looker}
\affiliation{ 
Sandia National Laboratories, Albuquerque, NM 87123-1106, USA
}%

\author{S. G. Patel}
\affiliation{ 
Sandia National Laboratories, Albuquerque, NM 87123-1106, USA
}%

\author{A. Porwitzky}
\affiliation{ 
Sandia National Laboratories, Albuquerque, NM 87123-1106, USA
}%

\author{G. A. Shipley}
\affiliation{ 
Sandia National Laboratories, Albuquerque, NM 87123-1106, USA
}%

\author{D. A. Uzdensky}
\affiliation{Center for Integrated Plasma Studies, Physics Department, UCB-390, University of Colorado, Boulder, Colorado, USA\looseness=-1 
}%

\author{D. A. Yager-Elorriaga}
\affiliation{ 
Sandia National Laboratories, Albuquerque, NM 87123-1106, USA
}%

\author{J.D. Hare *}%
\email{jdhare@mit.edu}
\affiliation{Plasma Science and Fusion Center, Massachusetts Institute of Technology, MA 02139, Cambridge, USA\looseness=-10000 
}%


\begin{abstract}
 We present evidence for strong radiative cooling in a pulsed-power-driven magnetic reconnection experiment. Two aluminum exploding wire arrays, driven by a 20 MA peak current, 300 ns rise time pulse from the Z machine (Sandia National Laboratories), generate strongly-driven plasma flows ($M_A \approx 7$) with anti-parallel magnetic fields, which form a reconnection layer ($S_L \approx 120$) at the mid-plane. The net cooling rate far exceeds the Alfvénic transit rate ($\tau_{\text{cool}}^{-1}/\tau_{\text{A}}^{-1} > 100$), leading to strong cooling of the reconnection layer. We determine the advected magnetic field and flow velocity using inductive probes positioned in the inflow to the layer, and inflow ion density and temperature from analysis of visible emission spectroscopy. A sharp decrease in X-ray emission from the reconnection layer, measured using filtered diodes and time-gated X-ray imaging, provides evidence for strong cooling of the reconnection layer after its initial formation. X-ray images also show localized hotspots, regions of strong X-ray emission, with velocities comparable to the expected outflow velocity from the reconnection layer. These hotspots are consistent with plasmoids observed in 3D radiative resistive magnetohydrodynamic simulations of the experiment. X-ray spectroscopy further indicates that the hotspots have a temperature ($\SI{170}{\electronvolt}$) much higher than the bulk layer ($\leq\SI{75}{\electronvolt}$) and inflow temperatures (about $\SI{2}{\electronvolt}$), and that these hotspots generate the majority of the high-energy ($>\SI{1}{\kilo\electronvolt}$) emission.

\end{abstract}

\maketitle

\newcommand{\ra}[1]{\renewcommand{\arraystretch}{#1}}

\setlength{\textfloatsep}{4pt plus 0.5pt minus 0.5pt}

\section{\label{sec:intro} Introduction}

Magnetic reconnection, a ubiquitous process in magnetized plasmas, is responsible for many highly energetic events in our Universe, such as solar flares and coronal mass ejections in our solar system, \citep{parker1963solar,masuda1994loop,yamada2010magnetic} reconnection events in the coronae of other stars, in the accretion disks and jets of Young Stellar Objects (YSOs), \citep{goodson1997time,feigelson1999high,benz2010physical} and in the interstellar medium \citep{zweibel1989magnetic,brandenburg1995effects,lazarian1999reconnection,heitsch2003fast}. Reconnection occurs when anti-parallel magnetic field lines advected by plasma flows generate a current sheet, also known as a reconnection layer. In the current sheet, the frozen-in flux condition breaks locally, allowing an abrupt reconfiguration of the magnetic field topology. \cite{parker1957sweet,yamada2010magnetic,ji2022magnetic} This process explosively converts magnetic energy into thermal and kinetic energy inside the reconnection layer.

Because of the dissipation of magnetic energy as heat, radiative emission is often the key, and sometimes the only, observable signature of reconnection in many astrophysical systems, such as in solar and YSO flares. \cite{somov1976physical,feigelson1999high,uzdensky2011magneticb,uzdensky2016radiative} Reconnection is also a proposed mechanism to explain the bursts of transient high-energy radiation observed from extreme relativistic astrophysical objects, such as black hole accretion disks and their coronae \citep{goodman2008reconnection,beloborodov2017radiative,werner2019particle,ripperda2020magnetic}, $\gamma$-ray bursts \citep{lyutikov2006electromagnetic,giannios2008prompt,zhang2010internal,uzdensky2011magnetic,mckinney2012reconnection}, and pulsar magnetospheres and winds\citep{lyubarsky2001reconnection,uzdensky2013physical,cerutti2015particle,cerutti2016modelling,philippov2018ab}. In astrophysical systems with strong reconnection-driven radiative emission, radiative cooling can be significant enough to rapidly remove internal energy from the system \citep{uzdensky2011magnetic,uzdensky2011magneticb,uzdensky2016radiative}. In particular, radiative cooling becomes important when the cooling rate $\tau_{\text{cool}}^{-1}$ dominates the Alfvén transit rate $\tau_A^{-1}$ in the reconnection layer, resulting in a large cooling parameter $R_{\text{cool}} \equiv \tau_{\text{cool}}^{-1}/\tau_A^{-1} \gg 1$. Here, $\tau_{\text{cool}} \equiv p / P_{\text{cool}}$ is the thermal pressure divided by the net volumetric cooling rate $P_{\text{cool}}$, and $\tau_A \equiv L/V_{A,in}$, where $L$ is the layer half-length, and $V_{A,in}$ is the Alfvén speed, calculated just outside the layer. In this regime, the plasma cools significantly before being ejected from the reconnection layer. 


The ubiquity of radiative emission in astrophysical plasmas has motivated theoretical and numerical investigation into radiatively-cooled magnetic reconnection. \citep{dorman1995one,uzdensky2011magnetic,uzdensky2016radiative,beloborodov2017radiative,jaroschek2009radiation, cerutti2014three, werner2019particle, mehlhaff2020kinetic,mehlhaff2021pair, schoeffler2019bright, schoeffler2023high, sironi2020kinetic, sridhar2021comptonization, sridhar2023comptonization, hakobyan2019effects, hakobyan2023radiative} Building on earlier work by Kulsrud and Dorman\cite{dorman1995one}, Uzdensky and McKinney\cite{uzdensky2011magnetic} presented the first theoretical description of radiatively-cooled Sweet-Parker-like reconnection. In this theory, the loss of internal energy via radiative emission cools the reconnection layer, leading to compression by the upstream magnetic pressure.
This in turn increases the radiative emission rate. When an increase in layer compression causes radiative losses to increase faster than Ohmic dissipation, rapid runaway compression and cooling of the reconnection layer occurs; this process is termed radiative collapse \cite{dorman1995one,uzdensky2011magnetic}. Uzdensky and McKinney\cite{uzdensky2011magnetic} showed that the Sweet-Parker reconnection rate is modified by a factor equal to the square root of the density compression ratio $A$, i.e. $E/V_{A,in}B_{in} \sim A^{1/2}S_L^{-1/2}$. Here, $E$ and $B_{in}$ are the reconnecting electric and magnetic fields respectively. The lower layer temperature results in a smaller (Spitzer) resistivity $\eta \sim T^{-3/2}$. Because of the strong compression $(A\gg1)$ and lower temperature (hence, lower Lundquist number $S_L = V_{A,in}L/\bar{\eta}$), this theory predicts an increased reconnection rate due to radiative collapse \cite{uzdensky2011magnetic}.

Numerical simulations of astrophysical systems in the non-relativistic resistive MHD regime show evidence for this cooling-driven compression of the reconnection layer \citep{dorman1995one,oreshina1998slow,laguna2017effect,ni2018magnetic,ni2018magnetic2}, consistent with theoretical predictions. More recently, this effect has also been observed in radiative-PIC (particle-in-cell) simulations of relativistic magnetic reconnection, which model reconnection physics in strongly-cooled extreme astrophysical systems exhibiting synchrotron and/or inverse external Compton cooling \citep{jaroschek2009radiation, cerutti2014three, werner2019particle, mehlhaff2020kinetic,mehlhaff2021pair, schoeffler2019bright, schoeffler2023high, sironi2020kinetic, sridhar2021comptonization, sridhar2023comptonization, hakobyan2019effects, hakobyan2023radiative}. Simulations of current sheets unstable to the plasmoid instability in electron-positron pair plasmas cooled via synchrotron emission have additionally shown cooling-driven compression of the density and reconnected magnetic flux inside magnetic islands or `plasmoids',\citep{loureiro2015magnetic,uzdensky2010fast} making them sites of enhanced radiative emission in the current sheet \citep{schoeffler2019bright,schoeffler2023high}.

Despite various numerical studies, experimental investigation of radiatively-cooled reconnection remains largely unexplored, in part due to the difficulty in achieving the cooling rates necessary for observing radiative collapse on experimental time scales. \, \citeauthor{yamada2010magnetic} provide a review of major laboratory experiments of magnetic reconnection\cite{yamada2010magnetic}. The earliest experimental validation of Sweet-Parker theory in a laboratory experiment was provided by the Magnetic Reconnection eXperiment (MRX), which generated a quasi-2D collision-dominated $S_L > 10^3$ reconnection layer in a toroidally symmetric geometry. \citep{yamada1997study,ji1999magnetic} MRX, and other magnetically-driven devices such as TREX \cite{olson2016experimental,olson2021regulation}, access a low-density magnetically-dominated regime $(n_e \sim10^{12}-\SI{e13}{\per \centi \meter \cubed}, \, T_e \sim 10 \, \text{eV}, \, \beta \ll 1)$ where radiative cooling is negligible. These experiments have provided significant insight into a variety of reconnection physics, such as evidence for strong ion heating \cite{hsu2000local}, two-fluid effects and Hall reconnection \cite{ren2005experimental,olson2016experimental,olson2021regulation}, as well as magnetic flux pile-up \citep{olson2021regulation}.

In contrast to low-$\beta$ magnetically-driven experiments, laser-driven experiments of magnetic reconnection provide access to a strongly-driven $\beta \gg 1$ high-energy-density (HED) regime $(n_e \sim \SI{e20}{\per \centi \meter \cubed}, \, T_e \sim 1000 \,\, \text{eV})$ \cite{rosenberg2015slowing,fox2012magnetic}. In these experiments, adjacent sub-millimeter spots on a solid target are irradiated with an intense Terawatt-class laser beam. The reconnecting magnetic fields are either self-generated by the Biermann battery effect \cite{rosenberg2015slowing,nilson2006magnetic,li2007observation,fox2021novel}, or supplied via external coils\cite{fiksel2014magnetic} or laser-driven capacitor coils\cite{chien2023non}. Laser-driven experiments have provided evidence for two-fluid effects,\cite{fiksel2014magnetic,rosenberg2015slowing,fiksel2014magnetic,fox2021novel} magnetic flux pile-up,\cite{fiksel2014magnetic} and particle acceleration in magnetic reconnection \cite{chien2023non}. However, despite the high operating pressure, the cooling parameter in these experiments was small as the plasma ions become fully-stripped at these high temperatures \cite{nilson2006magnetic,li2007observation,fox2012magnetic}, eliminating strong cooling by atomic transitions.

\begin{figure}[t!]
\includegraphics[page=1,width=0.48\textwidth]{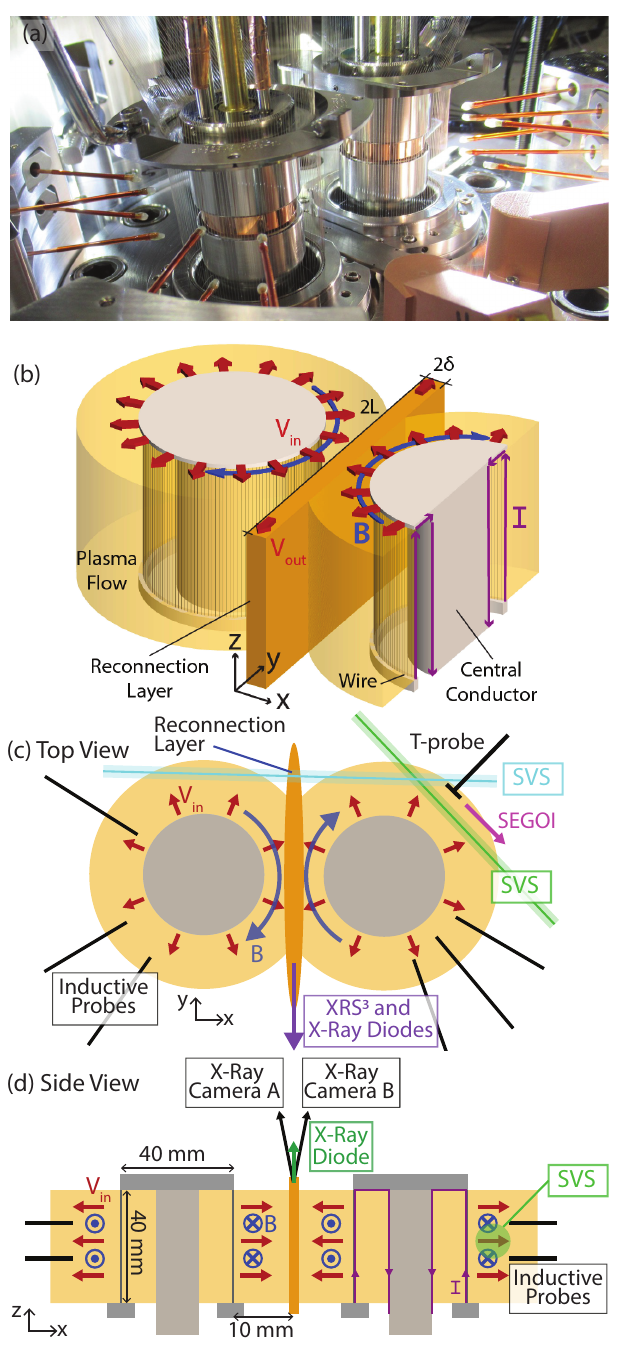}
\centering
\caption{\small (a) In-chamber image of the MARZ load hardware. (b) A 3D schematic of the MARZ load hardware, showing two exploding wire arrays, each with 150, \SI{75}{\micro \meter} diameter aluminum wires. Each array generates radially-diverging plasma flows (red) with frozen-in magnetic fields (blue), that generate a reconnection layer at the $x=0$ midplane. (c) Top ($xy$-plane) view of the load, showing the arrangement of inductive probes (black), and lines-of-sight of streaked visible spectroscopy (SVS) [green, blue], X-ray spectroscopy (XRS$^3$) and side-on X-ray diodes (purple), and the self-emission gated optical imager (SEGOI) [magneta]. (d) Side ($xz$-plane) view of the load, showing the lines-of-sight of the end-on X-ray diode (green), and the two X-ray cameras (black).}
\label{fig:load_setup}
\end{figure}

Pulsed-power-driven experiments are another class of strongly-driven $\beta \approx 0.1-1$ HED magnetic reconnection experiments \cite{lebedev2019exploring}. In these experiments, a strong ($\sim$~1~MA peak currrent peak currrent) time-varying ($100-\SI{250}{\nano\second}$) current pulse simultaneously drives two cylindrical exploding wire arrays placed side-by-side.\cite{suttle2016structure,hare2018experimental,hare2017anomalous,hare2017formation,suttle2019interactions} Each wire array generates radially-diverging (with respect to the array center) flows of magnetized plasma, which collide in the mid-plane, generating a reconnection layer. Experiments on the MAGPIE facility $(n_e \sim \SI{e18}{\per \centi \meter \cubed}, \, T_e \sim 50 \,\text{eV})$ using aluminum wires demonstrated cooling of the ions at a low $S_L <10$, measured via collective spatially-resolved optical Thomson scattering \cite{suttle2016structure,suttle2018ion}. Using lower-Z carbon wires, these experiments accessed higher Lundquist numbers $S_L \sim 100$ \cite{hare2017anomalous,hare2018experimental}, at which plasmoid formation was observed, unlike in the lower Lundquist number aluminum experiments \citep{suttle2016structure,suttle2018ion}. However, in these carbon experiments, there was negligible cooling of the reconnection layer, as the carbon ions were fully stripped. 

In this paper, we present results from the Magnetic Reconnection on Z (MARZ) experiments, which build on previous pulsed-power experiments and simultaneously demonstrate both a high $S_L \approx 120$ and a high cooling parameter $R_{\text{cool}} > 100$. The MARZ experimental platform generates a radiatively-cooled reconnection layer by driving a dual exploding wire array using the Z machine (20~MA peak current, \SI{300}{\nano\second} rise time, Sandia National Labs) \cite{sinars2020review}. The reconnection layer undergoes strong cooling, which is characterized by the rapid decline in X-ray emission generated from the layer. Furthermore, high-energy emission from the layer is dominated by localized fast-moving hotspots, which are consistent with magnetic islands produced by the plasmoid instability \cite{loureiro2015magnetic,uzdensky2010fast,schoeffler2023high}  which were seen in three-dimensional resistive magnetohydrodynamic simulations of the experiment.\citep{datta2024simulations} These experiments provide the first quantitative measurements of reconnection and plasmoid formation in a strongly radiatively-cooled regime, and directly characterize the high-energy radiative emission from the reconnection layer, using temporally- and spatially-resolved X-ray diagnostics. This, as mentioned earlier, is of particular astrophysical significance, because of the generation of high-energy emission in reconnecting astrophysical systems \cite{uzdensky2011magneticb,uzdensky2016radiative}.

Two- and three-dimensional radiative resistive magnetohydrodynamic (MHD) simulations of the MARZ experiments were previously reported in Ref.~\onlinecite{datta2024simulations}. The simulations were performed using GORGON --- an Eulerian resistive MHD code with van Leer advection \cite{chittenden2004equilibrium,ciardi2007evolution}. The simulations implemented both volumetric radiative loss and $P_{1/3}$ multi-group radiation transport, using spectral emissivity and opacity tables generated by the atomic code SpK \cite{crilly2022spk}. Line emission dominated in the simulations, providing strong cooling of the reconnection layer. Key results from these simulations are summarized as follows --- (1) strong radiative cooling ($R_{\text{cool}} \approx 100$) drove radiative collapse of the current sheet, resulting in decreased layer temperature and strong compression; and (2) the current sheet was unstable to the plasmoid instability, forming strongly-emitting plasmoids in the reconnection layer, which eventually collapsed due to radiative cooling \citep{datta2024simulations}. In this paper, we additionally compare our experimental findings with results from these simulations. Our previous paper \cite{datta2024plasmoid} provided details on the X-ray diagnostic results from these experiments; this present paper not only expands on the X-ray results, but also contributes results from additional diagnostics, providing context and further characterization of these experiments.  

This paper is structured as follows. In {\color{blue} \S}\ref{sec:hardware_diagnostics}, we describe the load hardware and the experimental diagnostics. Results from the experiments are described in {\color{blue} \S}\ref{sec:results}, while analysis of experimental data and discussion of key results, including radiative cooling and plasmoid formation, are provided in {\color{blue} \S}\ref{sec:discussion}. Finally, we outline key conclusions and future work in {\color{blue} \S}\ref{sec:conclusions}.

\section{Experimental and Diagnostic Setup}
\label{sec:hardware_diagnostics}

\subsection{MARZ Load Hardware}
\label{sec:hardware}

 \autoref{fig:load_setup}(a-b) show the load hardware. The load consists of two cylindrical exploding arrays, each with 150~equally-spaced, \SI{75}{\micro\metre} diameter aluminum wires. The array diameter is $\SI{40}{\milli \meter}$, and the array height is  $\SI{36}{\milli \meter}$. The center-to-center separation between the arrays is $\SI{60}{\milli \meter}$, giving a $\SI{10}{\milli\meter}$ distance between the wires and the mid-plane. Both wire arrays are over-massed, so they generate continuous plasma flows throughout the experiment without exploding \citep{lebedev2002snowplow,datta2023plasma}. A scaled experiment that matched the current per wire and driving magnetic field of the MARZ experiment was reported in Ref.~\onlinecite{datta2023plasma}, using a single planar wire array on the 1~MA COBRA machine. Results from these experiments show good ablation from \SI{75}{\micro\metre} diameter Al wires, with no closure of the inter-wire and the cathode-wire gaps \cite{datta2023plasma}.
 
 The Z machine \cite{sinars2020review} drives a 20 MA peak, \SI{300}{\nano\second} rise time current pulse through the load, which has an inductance of about $\SI{2.5}{\nano \henry}$. When current flows through the wires, the wires heat up resistively, and the wire material vaporizes and ionizes to create low-density coronal plasma surrounding the dense wire cores. Current density is concentrated within a thin skin region which generates coronal plasma around the stationary cores. The driving magnetic field points azimuthally inside the cathode-wire gap of each array, and rapidly drops to zero outside the array \citep{velikovich2002perfectly}. The global ${\bf j \times B}$ force, therefore, accelerates the coronal plasma radially outwards from each array, and the ablated plasma streams supersonically and super-Alfvénically into the vacuum region outside the arrays. The ablating plasma advects magnetic field from inside the cathode-wire gap to the outside, resulting in radially-diverging flows of magnetized plasma.\citep{lebedev2014formation} The plasma flows from each array advect frozen-in magnetic field to the mid-plane, where the field lines are anti-parallel and generate a reconnection layer (see \autoref{fig:load_setup}). 
 
 Three MARZ shots (MARZ1, MARZ2, and MARZ3) have been conducted on the Z machine so far. Each shot was fielded with identical load hardware and driving conditions, and an evolving set of diagnostics, as detailed in {\color{blue} \S}\ref{sec:diagnostic_setup}.

\begin{figure}[t!]
\includegraphics[page=2,width=0.5\textwidth]{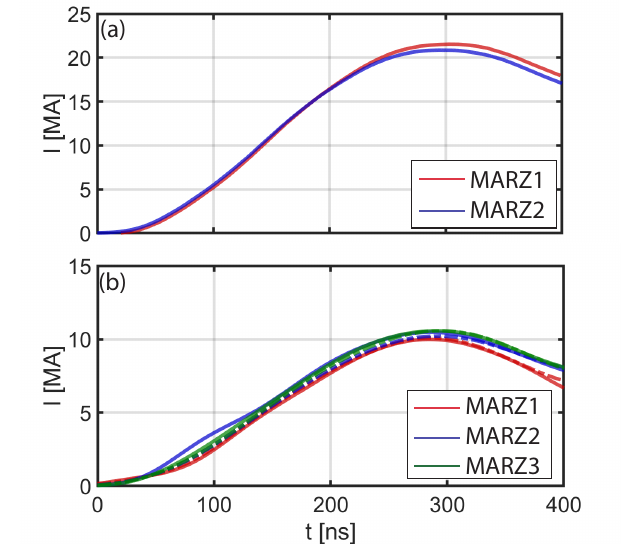}
\centering
\caption{\small (a) Averaged current measured by B-dot probes in the magnetically insulated transmission line (MITL) of the Z machine for two different shots (MARZ1 - red; MARZ2 - blue). The peak current is about 21 MA, and the rise time is about $\SI{300}{\nano \second}$. (b) Current measured by Photonic Doppler Velocimetry (PDV) in the east (solid) and west (dashed) arrays, showing equal current division between both arrays on all three shots. }
\label{fig:current_diagnostics}
\end{figure}
 
\subsection{Diagnostic Setup}
\label{sec:diagnostic_setup}

\autoref{fig:load_setup}(c-d) show the diagnostic setup. We categorize the diagnostics into current, inflow, and reconnection layer diagnostics. Current diagnostics, which include B-dot probes in the magnetically insulated transmission line (MITL) of the Z machine and Laser Photonic Doppler Velocimetry (PDV), monitor current delivery to the load. Inflow diagnostics, which include inductive probes and streaked visible spectroscopy, characterize the plasma ablating from the wires, which in turn, form the inflows into the reconnection layer. Finally, the reconnection layer diagnostics characterize the plasma in the current sheet, and consist of filtered X-ray diodes, X-ray imaging, and time-integrated X-ray spectroscopy. We provide more details on each diagnostic below.

\subsubsection{Current Diagnostics}

Dual-polarity B-dot probes in the MITL of the Z machine monitor the load current.\citep{webb2023radiation} These probes are calibrated, and their signals are numerically integrated to determine the current. PDV \cite{webb2023radiation,swanson2022development} is used to monitor the current delivered to each individual wire array. PDV tracks the velocity of a copper flyer plate which forms a section of the central conductor of each array, which accelerates due to the driving magnetic pressure. \cite{webb2023radiation} A comparison of the measured flyer plate velocity with 1D-MHD simulations is used to calculate the delivered current. Each array contains 4 separate PDV probes which record the flyer plate velocity at different azimuthal locations around the central conductor. The MITL B-dots and PDV are routinely fielded on the Z machine to characterize power flow; details of these systems are provided in Ref.~\onlinecite{webb2023radiation}.

\subsubsection{Inductive probes} 

\label{sec:probes}

 We position inductive probes at multiple radial locations around the wire arrays to measure the time- and space-resolved magnetic field advected by the plasma ablating from the wires. Each inductive probe consists of a single-turn loop created by connecting the inner conductor of a coaxial cable (\SI{2}{\milli \meter} outer diameter) to the outer conductor.\cite{byvank2017applied} In MARZ1, inductive probes were positioned at radial distances of \SI{5}{\milli \meter}, \SI{8}{\milli \meter}, \SI{11}{\milli \meter}, and \SI{14}{\milli \meter} from the wires, with the normals to the wire loops parallel to the azimuthal magnetic field. In MARZ2 and MARZ3, the probes were at \SI{5}{\milli \meter}, \SI{10}{\milli \meter}, and \SI{20}{\milli \meter}. The probes at different radii are at different azimuthal locations (see \autoref{fig:load_setup}c); however, the azimuthal variation in the measured magnetic field is expected to be small due to cylindrical symmetry \cite{lebedev2014formation,burdiak2017structure,datta2022structure,datta2022time}. We position two probes of opposite polarity at each location, separated vertically by \SI{1}{\centi \meter} (see \autoref{fig:load_setup}d). This allows us to eliminate the contribution of electrostatic voltages via common mode rejection. \cite{burdiak2017structure,datta2022time} Each probe is calibrated before use, and we numerically integrate the signals to determine the magnetic field.

\subsubsection{Streaked Visible Spectroscopy}

Streaked visible spectroscopy (SVS)\cite{webb2023radiation,schaeuble2021experimental} makes measurements of visible emission spectra from the plasma, along paths in the $xy$ plane, as shown in \autoref{fig:load_setup}c. Optical fibers collect and transmit light to a spectrometer (1 m McPherson Model 2061 scanning monochromator; 140~mm $\times$ 120~mm, 50 G/mm, $\SI{6563}{\angstrom}$ blaze diffraction grating) and a streak camera (Sydor streak camera; SI-800 CCD) setup. The tip of each fiber (Oz Optics, LPC-06-532-105/125-QM-0.8-1.81CL) consists of a MgF$_2$ anti-reflection coated collimating lens ($f = \SI{1.85}{\milli \meter}, \, \text{NA} = 0.22$). The beam divergence is $\SI{57}{\milli \radian}$, resulting in a spot diameter of about $\SI{5}{\milli \meter}$ at the center of the collection volume. The spectral range and resolution of the system are  $\SI{300}{\nano \meter}$ and $\SI{1.5}{\nano \meter}$ respectively. The sweep time is about $\SI{550}{\nano \second}$, and the temporal resolution is \SI{0.3}{\nano \second}. In MARZ1, we simultaneously record emission from the plasma ablating from the backside of the arrays at \SI{8}{\milli \meter} and \SI{17}{\milli \meter} from the wires (green line in \autoref{fig:load_setup}c), using separate SVS systems. In MARZ3, the visible spectroscopy line-of-sight (LOS) includes plasma ablating from both arrays and the plasma in the reconnection layer (blue line in \autoref{fig:load_setup}c), along $y = \SI{26.5}{\milli\meter}$ and $y = \SI{34}{\milli\meter}$.

\subsubsection{Visible Self-Emission Imaging}

In MARZ1, we positioned an additional inductive probe with a $10$ mm long \SI{1}{\milli\meter} diameter glass rod attached to its tip, at \SI{15}{\milli \meter} from the wires (hereafter, referred to as the `T-probe') (see \autoref{fig:load_setup}c). It provides an extended obstacle that creates a detached bow shock when the flow interacts with the probe.  We observe the bow shock using a self-emission gated optical imager (SEGOI) \cite{webb2023radiation}. SEGOI is an 8-frame camera that records 2D self-emission images in the visible range (540-\SI{650}{\nano \meter}) on 8 separate micro-channel plates (MCPs). We record images between 320-\SI{367}{\nano \second}, with a \SI{7}{\nano \second} inter-frame time, \SI{1}{\nano \second} exposure, and a $\SI{8}{\milli\meter}$ diameter field of view. SEGOI also captures a 1D streak image (sweep time = 300-\SI{400}{\nano\second}) along a line  parallel to the T-probe axis, \SI{2}{\milli \meter} below the probe. 


\begin{figure}[t!]
\includegraphics[page=3,width=0.48\textwidth]{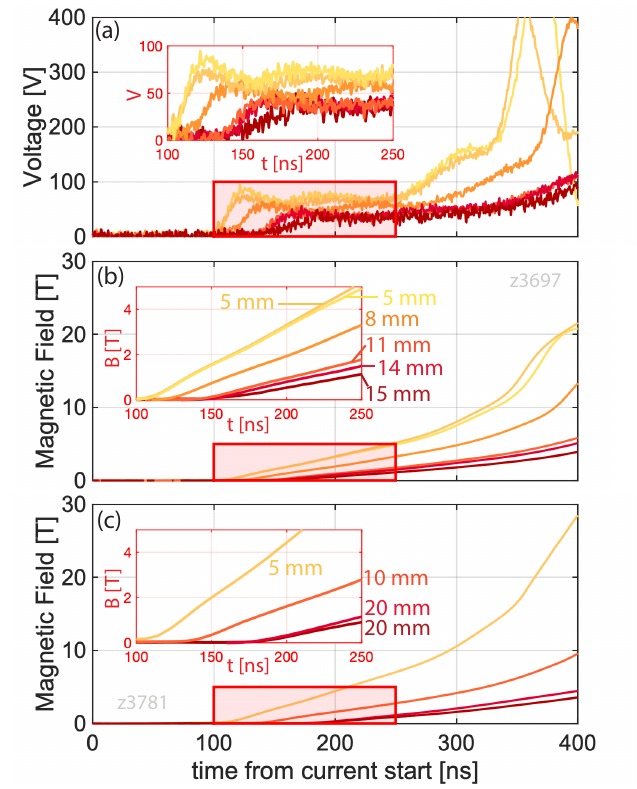}
\centering
\caption{\small (a) Inductive voltage signals from probes placed at radial distances 5, 8, 11, 14, and 15 mm from the wires in MARZ1. Signals are delayed with respect to one another due to the transit time of the magnetic field advected by the plasma between the probe locations. (b) Time-resolved magnetic field measurements at different radii around the array in MARZ1. (c) Magnetic field measurements in MARZ3.}
\label{fig:inductive_probes}
\end{figure}

 \subsubsection{X-Ray Diodes}

Silicon diodes \cite{webb2023radiation} record the X-ray power generated from the reconnection layer. In MARZ1 and MARZ2, the diode viewed the reconnection layer from the side (side-on). The diode was filtered with $\SI{2}{\micro\meter}$ of aluminized Mylar (\autoref{fig:load_setup}c). The $T=0.5$ transmission cut-off for this filter is at about \SI{100}{\electronvolt}. In MARZ3, the diode viewed the reconnection layer from the top (end-on) [\autoref{fig:load_setup}d], and was filtered with $\SI{8}{\micro \meter}$-thick beryllium. The $T=0.5$ cut-off is at roughly $\SI{1}{\kilo \electronvolt}$. Each diode has a \SI{22.5}{\micro \meter} active layer and a nominal \SI{0.6}{\milli \meter} diameter. The diode response is 0.276 A/W.

\begin{figure*}[t!]
\includegraphics[page=4,width=1\textwidth]{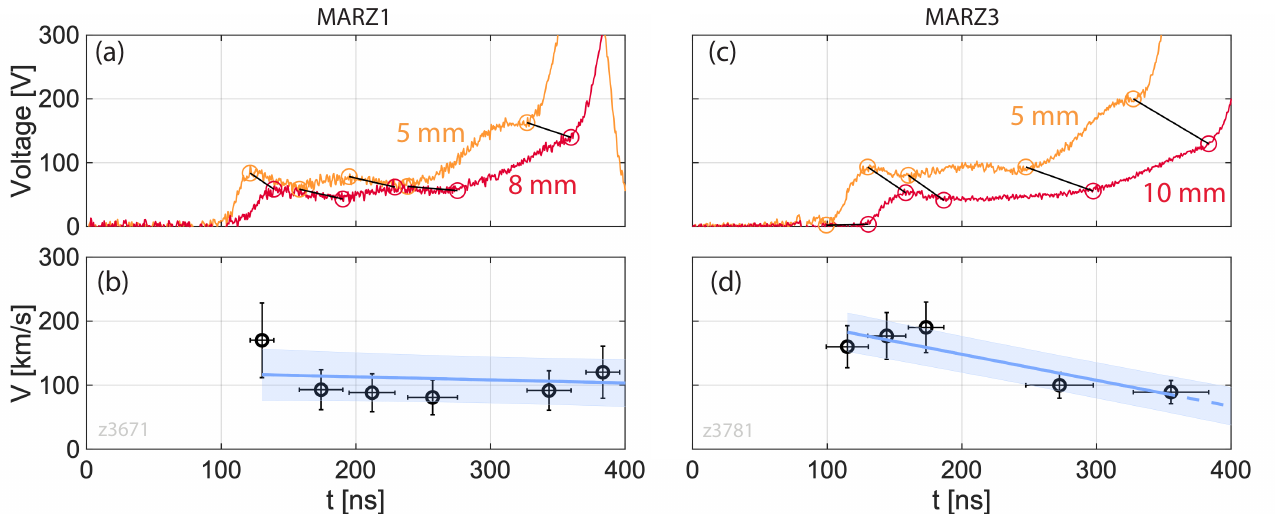}
\centering
\caption{\small (a) Inductive voltage signals from probes placed at \SI{5}{\milli \meter} and \SI{8}{\milli \meter} from the wires in MARZ1, showing delay between the signals. (b) Estimate of the average flow velocity between 5-\SI{8}{\milli \meter} from the time delay of the inductive probe signals in MARZ1. (c) Inductive voltage signals from probes placed at \SI{5}{\milli \meter} and \SI{10}{\milli \meter} from the wires in MARZ3. (d) Estimate of the average flow velocity between 5-\SI{10}{\milli \meter} from the time delay of the inductive probe signals in MARZ3. }
\label{fig:velocity}
\end{figure*}

\subsubsection{X-Ray Imaging}

We image the reconnection layer using two time-gated ultra-fast X-ray imaging (UXI) cameras \cite{webb2023radiation}. The cameras provide a $25 \times \SI{12.5}{\milli \meter \squared}$ ($1025 \text{px} \times 512 \text{px}$) field of view through a $\SI{500}{\micro \meter}$ diameter pinhole (magnification = $1\times$, geometric resolution $\approx \SI{1}{\milli \meter}$). The cameras view the reconnection layer with polar angles of $\theta= \SI{9}{\degree}$ and $\theta = \SI{12}{\degree}$ with respect to the z-axis (see \autoref{fig:load_setup}d), and with azimuthal angles (from the $x-$axis) of $ \phi = \SI{170}{\degree}$ and $\phi = \SI{40}{\degree}$ (not shown in \autoref{fig:load_setup}), thus viewing both the top and side of the layer. The pinholes are filtered with $\SI{2}{\micro \meter}$ thick aluminized Mylar, which filters out photons with energies $<\SI{100}{\electronvolt}$. Each camera records 4 images with a \SI{20}{\nano \second} inter-frame time, and a \SI{10}{\nano \second} exposure time. Data from this diagnostic is only available for MARZ3.

In addition to time-gated imaging, we record time-integrated X-ray images of the reconnection layer, using two pinhole cameras viewing the layer from the top with polar angles of roughly $\SI{5}{\degree}$. Pinhole diameters of \SI{300}{\micro\meter} and \SI{500}{\micro\meter} are used. Each camera has 3 pinholes of the same diameter but different filtration that generate three images (magnification $\approx 0.5$, resolution $\approx 450-\SI{750}{\micro\meter}$) of the reconnection layer on a \SI{64}{\milli\meter} $\times$ \SI{34}{\milli\meter} image plate.


 \subsubsection{Time-integrated X-Ray Spectroscopy}

An X-ray scattering spectrometer (XRS$^3$) \cite{harding2015analysis,webb2023radiation} with a spherically-bent quartz crystal provides time-integrated spatially-resolved (along the out-of-plane $z$ direction, resolution: $\Delta z \approx \SI{200}{\micro \meter}$) measurements of X-ray emission spectra from the reconnection layer. The range and spectral resolution of the spectrometer in the MARZ experiments were 1.5-$\SI{1.9}{\kilo \electronvolt}$ and $\Delta E \approx \SI{0.5}{\electronvolt}$ respectively. We record the X-ray spectrum on an image plate (Fuji TR), filtered with a $\SI{11}{\micro \meter}$ thick beryllium filter. Data was recorded in either of two configurations --- (1) \SI{150}{\milli \meter} radius crystal, crystal-to-target separation = $\SI{800}{\milli \meter}$, and a $\SI{8}{\micro \meter}$ kapton filter on the spectrometer entrance slit; and (2) \SI{200}{\milli \meter} radius crystal, crystal-to-target separation = $\SI{500}{\milli \meter}$, and no kapton filter. Configuration 1 was used for MARZ1 and MARZ2, while configuration 2 was used for MARZ3.

\section{Results}
\label{sec:results}

We describe the experimental results from our diagnostics in this section. In MARZ2, damage to one of the arrays during installation produced results which were unreproducible. Therefore, we show results primarily from MARZ1 and MARZ3, with some exceptions.

\subsection{Current Measurements}

\label{sec:current_measurements}

\autoref{fig:current_diagnostics}a shows the averaged current measured by the MITL B-dot probes in MARZ1 and MARZ2. The Z machine consistently delivered a peak current of roughly 21~MA, with a rise time of about \SI{300}{\nano\second} across the two shots. The shot-to-shot variation in the delivered current was $< 5\%$.

\autoref{fig:current_diagnostics}b shows the current measured by the PDV diagnostic for all three shots. We show the averaged current for the east (solid line) and west (dashed line) arrays in each shot. \autoref{fig:current_diagnostics}b shows equal current division between the arrays. As expected, the shape of the current pulse measured by PDV matches that of the current measured by the MITL B-dots. The peak value in each array is roughly 10~MA, showing negligible current loss between the MITL and the load. Current measurements by the MITL B-dot probes are not available for MARZ3, but the PDV measurements in \autoref{fig:current_diagnostics}b show that the current delivered to the load in this shot was consistent with the other two shots.

\begin{figure*}[t!]
\includegraphics[page=5,width=1\textwidth]{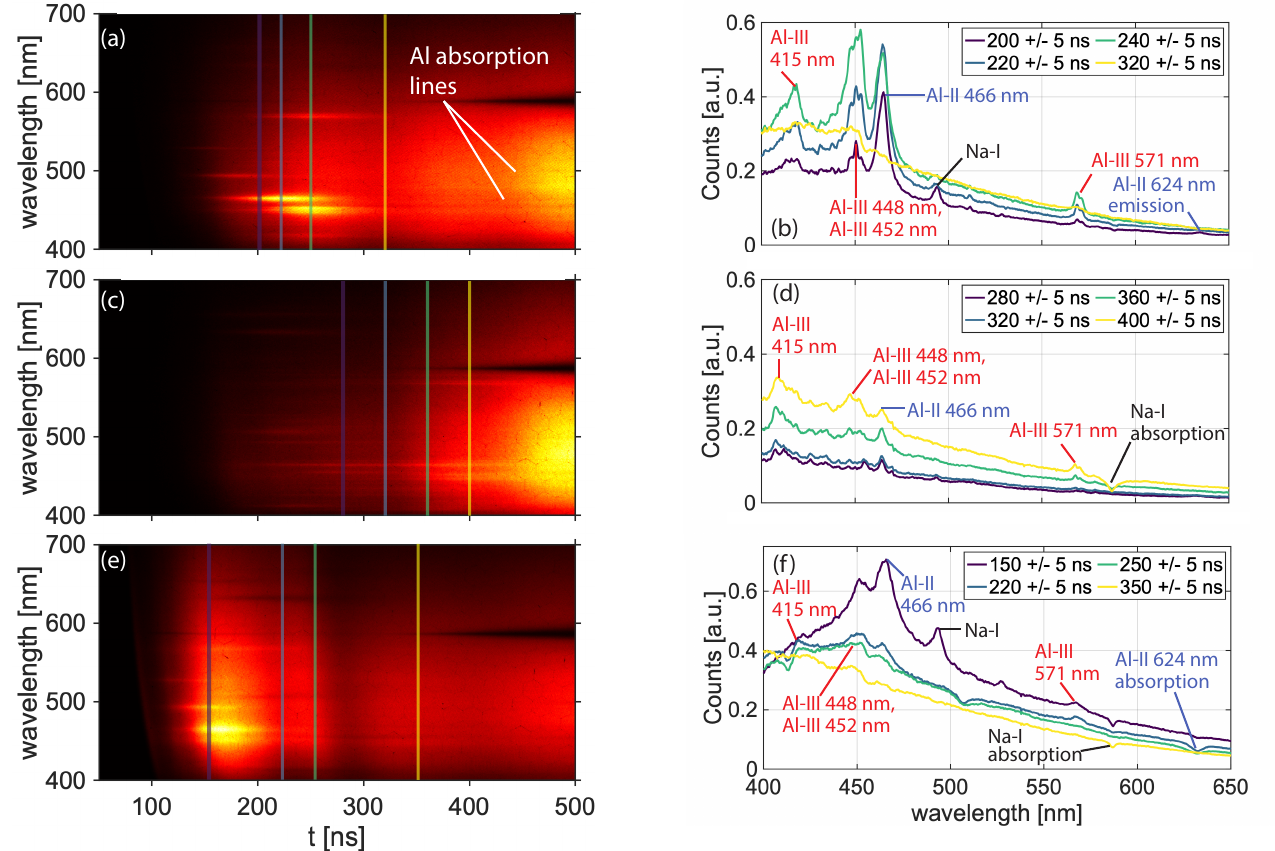}
\centering
\caption{\small  Streaked visible spectrum recorded (a) $\SI{8}{\milli \meter}$ from the wires and (c) $\SI{17}{\milli \meter}$ from the wires in MARZ1. (b) The visible spectrum at $t = 200, \, 220, \, 240\,\, \& \,\SI{320}{\nano \second}$ at \SI{8}{\milli\meter} from the wires. (d) The visible spectrum at $t = 280, \, 320, \, 360\,\, \& \, \SI{400}{\nano \second}$ at \SI{17}{\milli\meter} from the wires. These spectra are averaged over \SI{10}{\nano \second}. Identified emission lines are shown in (b) and (d). (e) Streaked visible spectrum in MARZ3. The diagnostic line of sight includes plasma in the reconnection layer and that ablating from the wires. (f) The visible spectrum at $t = 150, \, 220, \, 250\,\, \& \, \SI{350}{\nano \second}$ averaged over 10~ns. The Al-II~624~nm line appears as an absorption feature. }
\label{fig:SVS}
\end{figure*}

\subsection{Magnetic Field and Velocity in the Inflow Region}
\label{sec:magnetic_filed_results}

\autoref{fig:inductive_probes}a shows the voltage signals from probes in MARZ1. We only show the inductive component of the signals $\bar{V} = 0.5(V_+ - V_-)$, determined from common mode rejection of signals from the two opposite-polarity probes at each location. The signals in \autoref{fig:inductive_probes}a are all similar in shape, but displaced in time, which is expected due to the advection of the frozen-in magnetic field by the plasma between the locations of the probes. \cite{lebedev2014formation,datta2022time} We note that the probes are placed on the side of the arrays opposite to that of the reconnection layer (as shown in \autoref{fig:load_setup}), so they measure the ‘unperturbed’ magnetic field in the inflow, not affected by reconnection. 

We numerically integrate the signals in \autoref{fig:inductive_probes}a to determine the magnetic field, as shown in \autoref{fig:inductive_probes}b. The advected magnetic field increases with time, due to a rise in the driving current, and decreases with distance from the wires, consistent with previous measurements in exploding wire arrays.\cite{lebedev2014formation,burdiak2017structure,datta2022structure,datta2022time} In \autoref{fig:inductive_probes}b, two separate probes at $\SI{5}{\milli \meter}$ from the wires record the magnetic field from different arrays. Both probes exhibit similar magnetic fields, consistent with equal current splitting between the arrays, as described in {\color{blue} \S}\ref{sec:current_measurements}.

\autoref{fig:inductive_probes}c shows the magnetic field recorded in MARZ3. The recorded magnetic field is slightly larger than that in MARZ1. Here, inductive probes measuring the magnetic field from the two arrays at \SI{20}{\milli \meter} from the wires show identical magnetic fields, consistent with equal current splitting. We note that in MARZ3, the two opposite-polarity probes at 5~mm recorded significantly different signals. The signal on the bottom probe was consistent with the expected field magnitude (based on measurements in MARZ1), while the other was anomalously high. Therefore, we discard the anomalous signal and only report data from one probe at 5~mm in \autoref{fig:inductive_probes}c.

Finally, we note that probes measuring the advected magnetic field from different arrays at the same radial distance in MARZ2 measured significantly higher advected magnetic field on the west array, despite equal current splitting observed in \autoref{fig:current_diagnostics}b. The west array was partly damaged during installation on this shot, and the probes were measuring the magnetic field from the damaged section of the array.

The delay in the voltage signals between probes at different locations provides an estimate of the average flow velocity.\cite{datta2022time} \autoref{fig:velocity}a shows the voltage signals recorded by probes in MARZ1 at \SI{5}{\milli \meter} and \SI{8}{\milli \meter} respectively. The signals show several identifiable features, indicated via circles in \autoref{fig:velocity}a. By tracking the transit time of these features, we estimate the flow velocity, as shown in \autoref{fig:velocity}b. The flow velocity is roughly \SI{110}{\kilo \meter \per \second}, consistent with flow velocities previously recorded in pulsed-power-driven wire arrays \cite{lebedev2019exploring,suttle2019interactions}. In \autoref{fig:velocity}c, we show the inductive probe voltage measurements at \SI{5}{\milli \meter} and \SI{10}{\milli \meter} from the wires for MARZ3, together with the estimated flow velocity in \autoref{fig:velocity}d. On this shot, the recorded flow velocity varied between $100-\SI{200}{\kilo\meter\per\second}$.

\subsection{Measurements of Visible Spectra in the Inflow}
\label{sec:svs_results}

\autoref{fig:SVS} shows streak images of the visible emission spectra collected at \SI{8}{\milli \meter} and \SI{17}{\milli \meter} from the wires respectively in MARZ1. Note that these spectra are from the side of the wire arrays opposite the reconnection layer (see \autoref{fig:load_setup}c). We have applied corrections to these spectra for distortions by streak camera optics, timing corrections for spectral differences in photon transit delay over the fiber length, as well as relative intensity corrections due to the wavelength-dependent response of the spectrometer \cite{schaeuble2021experimental}. Wavelength calibration and instrument broadening (about \SI{1.5}{\nano\meter}) were determined using preshot images of \SI{458}{\nano\meter} and \SI{543}{\nano\meter} laser lines recorded by each SVS system. 

The streak camera first records emission at roughly \SI{90}{\nano \second} for the \SI{8}{\milli \meter} system (\autoref{fig:SVS}a).  This corresponds to an average flow velocity of roughly \SI{90}{\kilo \meter \per \second} between the wire and probe locations; the estimated velocity is consistent with that estimated from inductive probe measurements (\autoref{fig:velocity}). Compared to the spectra at \SI{8}{\milli \meter}, emission at \SI{17}{\milli \meter} is first recorded later at roughly \SI{140}{\nano\second}, corresponding to an average velocity of roughly \SI{120}{\kilo\meter\per\second} (\autoref{fig:SVS}c).  Finally, by integrating the spectral intensity in wavelength space, we find that the temporal evolution of intensity $\int I(\lambda) d \lambda$ (which depends on the plasma density) roughly matches that of the advected magnetic field shown in \autoref{fig:inductive_probes}. 

\autoref{fig:SVS}(b and d) show lineouts of the streak images at different times, each averaged over $\SI{10}{\nano \second}$. The spectra at both radial locations show well-defined Al-II and Al-III emission lines, which correspond to transitions in singly- and doubly-ionized aluminum respectively. Later in time, continuum emission begins to dominate over line emission. This occurs at around \SI{300}{\nano \second} and \SI{450}{\nano \second} for the \SI{8}{\milli \meter} and \SI{17}{\milli \meter} observations respectively. Late in time, the spectra also exhibit absorption features corresponding to Al-II and Al-III transitions, as indicated in \autoref{fig:SVS}. We use the Al emission lines to estimate time-resolved values of ion density and electron temperature in the inflow region. This analysis, performed using collisonal-radiative and radiation transport modeling, is described in \S\ref{sec:svs_snalaysis}.

In MARZ3, the diagnostic LOS included plasma in the reconnection layer and that ablating from both wire arrays, as shown in \autoref{fig:load_setup}c (blue line). The spectra, shown in \autoref{fig:SVS}(e-f), are similar to that recorded in MARZ1 (\autoref{fig:SVS}), with well-defined Al-II and Al-III emission features. However, between 150-350~ns,  the Al-II~624~nm transition, which appears as an emission line in \autoref{fig:SVS}(a-d), now appears as an absorption feature in \autoref{fig:SVS}(e-f). We discuss the origin of this absorption feature in \S\ref{sec:svs_snalaysis}.

In addition, the spectra exhibit features associated with the coatings on the optical components. The streak images show a stray Na-I line at roughly \SI{485}{\nano \meter} early in time, as well as a strong Na-I absorption feature later in time. These features are not generated by the aluminum plasma, but are instead stray features generated from the optical coatings in the system. These stray features appear at around the same time in all streak images. 


\begin{figure}[t!]
\includegraphics[page=9,width=0.45\textwidth]{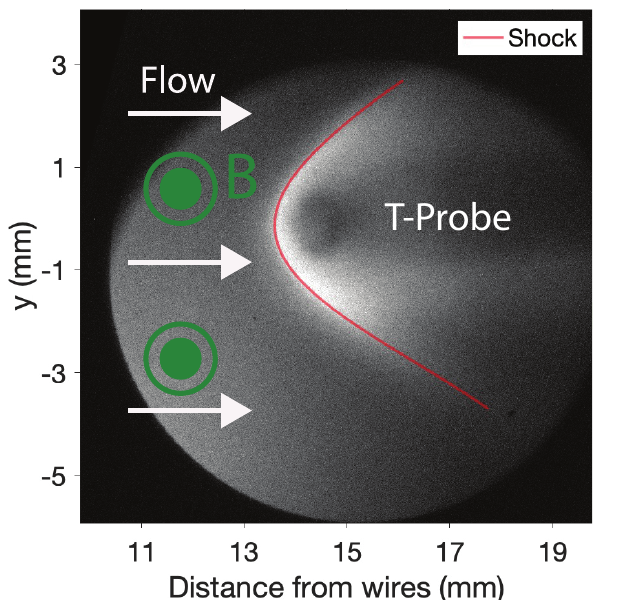}
\centering
\caption{\small Gated optical self-emission image of a bow shock around the T-probe. The bow shock exhibits a Mach angle of about $\SI{30}{\degree}$.
 }
\label{fig:bow_shock}
\end{figure}

\subsection{Bow shock Imaging}
\label{sec:bow_shock_results}

\autoref{fig:bow_shock} shows an optical self-emission image of the T-bar probe at \SI{346}{\nano \second}. A bow shock, which appears as a curved region of enhanced emission, forms around the T-probe. Shock formation around the T-probe provides visual confirmation of wire array ablation and generation of supersonic flow. Multi-frame self-emission images, as well as the 1D streak image of the bow shock, show that the shock front remains invariant in time between 300-\SI{400}{\nano \second}. The shock angle, determined from the derivative of the shock front position (red curve in \autoref{fig:bow_shock}),  asymptotes to about \SI{30}{\degree}.

\subsection{X-Ray emission from the reconnection layer}

 We now present results from the reconnection layer diagnostics, which characterize the temporal, spatial, and spectral properties of X-ray emission from the reconnection layer.

\begin{figure}[b!]
\includegraphics[page=7,width=0.5\textwidth]{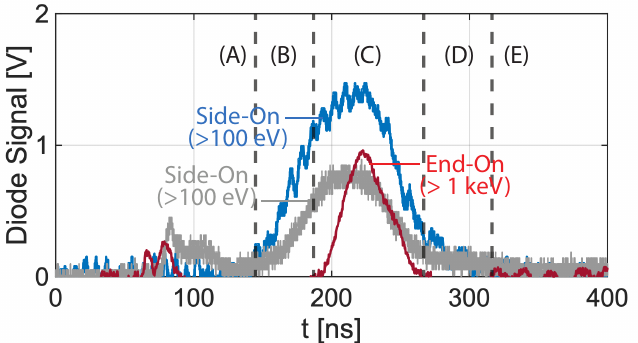}
\centering
\caption{\small X-ray power emitted from the reconnection layer as measured by the side-on (blue, gray) and end-on diodes (red).
 }
\label{fig:diodes}
\end{figure}

\subsubsection{X-Ray diodes}
\label{sec:diodes_results}

The X-ray diodes characterize the temporal evolution of X-ray emission from the reconnection layer. \autoref{fig:diodes} shows the signals from the side-on (blue, grey) and end-on (red) diodes. All three diode signals exhibit a peak in X-ray emission at about \SI{220}{\nano \second}. For the side-on diodes, which measure $>\SI{100}{\electronvolt}$ photons, the emission first ramps up around \SI{150}{\nano \second}, and the full-width-at-half-maximum (FWHM) of the signal is about $80-\SI{90}{\nano\second}$. Similarly, the signal from the end-on diode, which records comparatively harder X-rays with energy $>\SI{1}{\kilo \electronvolt}$,  initially ramps up around \SI{200}{\nano \second}, and exhibits a FWHM of roughly $\SI{50}{\nano \second}$. The X-ray emission peaks around \SI{220}{\nano \second} on all diodes, and then falls sharply. The shape of the X-ray emission is much sharper than that of the driving current pulse, which peaks about $\SI{100}{\nano \second}$ after the peak in X-ray emission (see \autoref{fig:current_diagnostics}). This shows that the emission feature is related to the dynamics of the current sheet, rather than the driving current. In addition, the diodes consistently record a small emission feature at about $\SI{100}{\nano\second}$. This feature may be related to the initial arrival of plasma at the mid-plane.

\begin{figure*}[t!]
\includegraphics[page=8,width=1\textwidth]{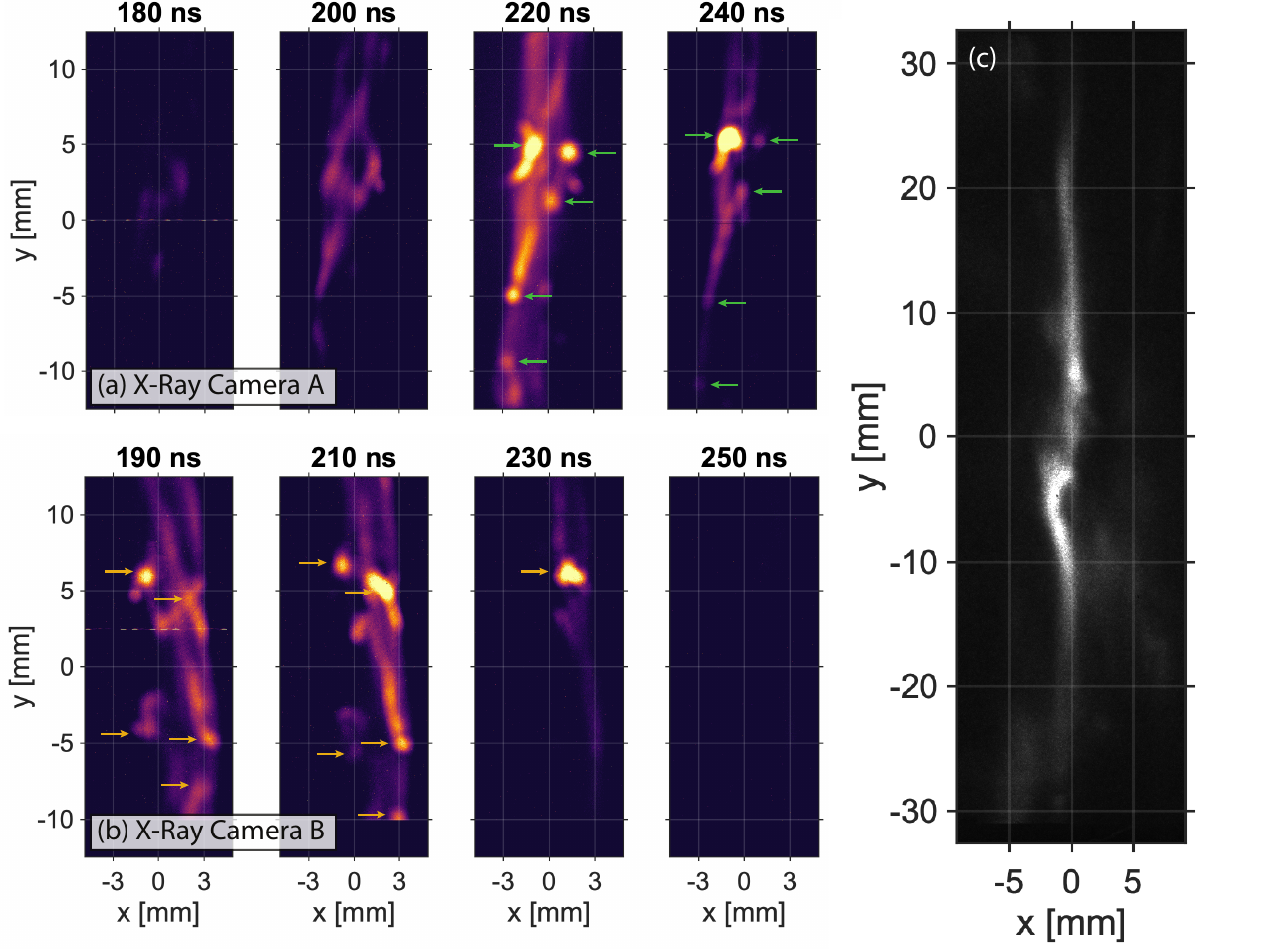}
\centering
\caption{\small  (a-b) Time-gated X-ray images ($\SI{10}{\nano \second}$ exposure time) of the reconnection layer at between 180-\SI{250}{\nano\second}, recorded using camera A (top row) and camera B (bottom row). X-ray images show brightly emitting hotspots (green and yellow arrows) embedded in an enlonged layer. (c) Time-integrated X-ray image of the reconnection layer. The image was recorded with a \SI{300}{\micro\meter} diameter pinhole, filtered with $\SI{2}{\micro\meter}$ aluminized Mylar.
 }
\label{fig:XrayImages}
\end{figure*}

\subsubsection{Time-gated X-ray Imaging}
\label{sec:uxi_results}

\autoref{fig:XrayImages}(a-b) shows time-gated images of the reconnection layer. Camera A ($\theta = \SI{9}{\degree}, \, \phi = \SI{170}{\degree}$) recorded images between 190-\SI{250}{\nano \second} at $\SI{20}{\nano\second}$ intervals, while camera B ($\theta = \SI{12}{\degree}, \, \phi = \SI{40}{\degree}$) recorded images between 180-\SI{240}{\nano \second}, again at $\SI{20}{\nano\second}$ intervals. Images from both cameras show an elongated layer of bright emission. The intensity of ($> 
\SI{100}{\electronvolt}$) X-ray emission increases initially, consistent with the formation of the reconnection layer, and falls thereafter. Along with the brightness of the emission, the width of the emitting region (along $x$) also decreases with time. Peak emission is recorded between $210-\SI{220}{\nano \second}$. By \SI{250}{\nano \second}, the emission has fallen significantly, and the layer is no longer visible on the X-ray cameras.

The X-ray images provide information about the spatial distribution of emission from the reconnection layer. Emission is highly inhomogeneous ---  \autoref{fig:XrayImages}(a-b) shows sub-millimeter scale regions of enhanced emission embedded within the less brightly emitting layer. The intensity of emission from the hotspots is $>10$ times higher than the average intensity from the rest of the layer. The presence of emission hotspots indicates localized regions of plasma with higher temperature or density relative to the rest of the layer. 

\begin{figure}[t!]
\includegraphics[page=10,width=0.5\textwidth]{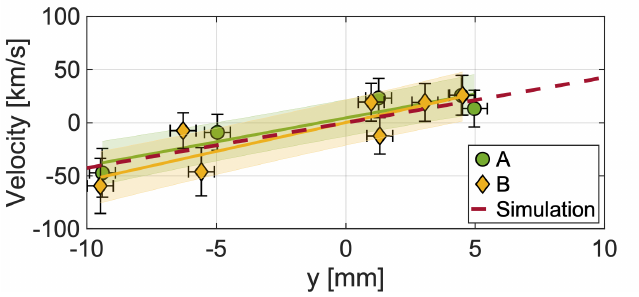}
\centering
\caption{\small Hotspot velocity estimated from the translation of the hotspots in the X-ray images of the reconnection layer. Green circles show velocity estimated from Camera A (220-240 ns), while yellow diamonds are for Camera B (190-210 ns). Solid lines show linear fits to the data. Dashed red line is the simulation outflow velocity from the reconnection layer.
 }
\label{fig:UXI_velocity}
\end{figure}

These hotspots, indicated via green and yellow arrows in \autoref{fig:XrayImages}(a-b), can be observed in images from both cameras to travel away from the center of the layer. We track the translation of the hotspot centroids between successive frames to estimate their velocities. \autoref{fig:UXI_velocity} shows the estimated hotspot velocity calculated using images from camera A between 220-\SI{240}{\nano \second} (green), and from camera B between 190-\SI{210}{\nano \second} (yellow). The hotspot velocities are consistent between both cameras. Hotspots accelerate along the $\pm y$-direction, away from the center of the layer. Hotspot velocity increases from \SI{0}{\kilo \meter \per \second} to about \SI{50}{\kilo \meter \per \second} over a distance of roughly \SI{10}{\milli \meter}. We will show in {\color{blue} \S}\ref{sec:cooling_rate} that the observed hotspot velocity is consistent with the expected velocity of the outflows from the reconnection layer.

\subsubsection{Time-integrated X-ray Imaging}
\label{sec:ATIPC_results}


\autoref{fig:XrayImages}c shows a time-integrated X-ray image of the reconnection layer. The image is recorded with a \SI{300}{\micro\meter} diameter pinhole, filtered with $\SI{2}{\micro\meter}$ aluminized Mylar, identical to the time-gated X-ray cameras in {\color{blue} \S}\ref{sec:uxi_results}. \autoref{fig:XrayImages}c shows an elongated region of bright emission. The extent of the emission region in the $y-$direction is about \SI{60}{\milli\meter}, while the FWHM along the $x-$direction is $1.6\pm\SI{0.5}{\milli\meter}$. The FWHM of the emitting region was determined by fitting a Gaussian function to the intensity variation along $x$ at different $y-$positions. Here, we only show one of the recorded time-integrated X-ray images; however, the features of the image in \autoref{fig:XrayImages}c are consistent with the other images recorded in the experiment.


\subsubsection{X-ray spectroscopy}
\label{sec:xrs3_results}

\begin{figure*}[t!]
\includegraphics[page=11,width=1\textwidth]{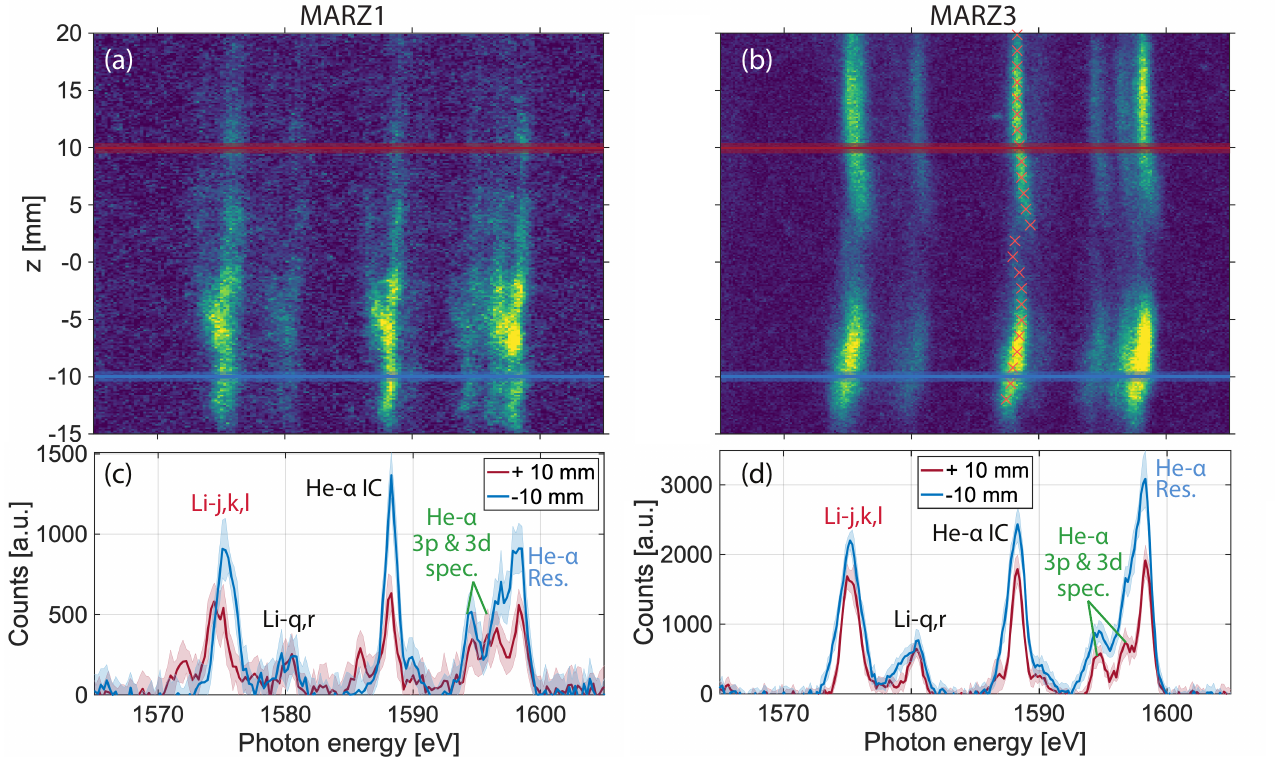}
\centering
\caption{\small (a-b) Time-integrated X-ray emission spectra recorded in MARZ1 and MARZ3. (c-d) Lineouts of the X-ray spectra at $z = \SI{10}{\milli\meter}$ (red) and $z = \SI{-10}{\milli\meter}$ (blue), showing Al K-shell emission lines. These include the He-$\alpha$ resonance and inter-combination (IC) lines, the He-$\alpha$ transitions with spectator electrons, and the Li-like satellite transitions. Shaded regions represent the standard deviation of the spectra inside the integration window.
 }
\label{fig:Xray_spectra}
\end{figure*}

\autoref{fig:Xray_spectra}(a-b) show the time-integrated spectrum of the X-ray emission from the reconnection layer for MARZ1 and MARZ3. Emission lines with energies 1570-\SI{1600}{\electronvolt} were observed in both shots. Although the output is time-integrated, the end-on diode signal (filtered with \SI{8}{\micro \meter} Be), which measured $>\SI{1}{\kilo\electronvolt}$ X-ray emission (see \autoref{fig:diodes}), shows that the spectrum was generated around $220 \pm 25$ ns. 

Lineouts of the recorded spectrum averaged over $z = 10 \pm \SI{0.5}{\milli \meter}$ and $z = -10 \pm \SI{0.5}{\milli \meter}$ are shown in \autoref{fig:Xray_spectra}(c-d). We label the Al K-shell emission lines, which include He-like and Li-like satellite transitions. The He-like lines correspond to transitions in Al-XII ions (2 bound electrons, $Z = 11$). Identified He-like lines include the He-$\alpha$ resonance line (\SI{1598}{\electronvolt}), the He-$\alpha$ inter-combination (IC) line (\SI{1588}{\electronvolt}), and He-$\alpha$ resonance lines with 3p and 3d spectator electrons (\SI{1594}{\electronvolt}, \SI{1596}{\electronvolt}). The He-$\alpha$ resonance line ($2p^{1}P_1 \rightarrow 1s^{1}S_0$) represents a transition to the ground state $^{1}S_0$ from the next highest energy state $^{1}P_{1}$ of the singlet system, while the inter-combination transition ($2p^{3}P_1 \rightarrow 1s^{1}S_0$) occurs between the upper term of the triplet system $^{3}P_1$ and the lower term of the singlet system $^{1}S_0$ \citep{herzberg1944atomic,sobelman2012atomic}. The transitions shielded by spectator electrons appear at energies lower than the resonance transition. In the recorded spectra, the He-$\alpha$ resonance and IC lines exhibit roughly similar intensities, while the He-$\alpha$ transitions with spectator electrons have lower intensities.

The spectra additionally exhibit Li-like satellite lines. Satellites are transitions from doubly excited states, typically populated via dielectronic recombination \citep{sobelman2012atomic}. The Li-j (\SI{1574.2}{\electronvolt}), Li-k (\SI{1575.0}{\electronvolt}), and Li-l (\SI{1574.3}{\electronvolt}) lines have similar upper ($1s2p^{2}$) and lower ($1s^{2}2p^{1}$) electronic configurations. However, they represent transitions between different combinations of the upper $D_{5/2,3/2}$ and lower $P_{3/2,1/2}$ terms in the doublet system, i.e. these are transitions from upper to lower states with the same orbital angular momentum ($L = 2 \rightarrow L = 1$) but different total angular momentum ($\text{j}: D_{5/2} \rightarrow P_{3/2}, \, \text{k}: D_{3/2} \rightarrow P_{1/2}, \, \text{l}: D_{3/2} \rightarrow P_{3/2}$) \citep{herzberg1944atomic,sobelman2012atomic}. The Li-j,k,l satellites thus have similar energies, and we do not resolve them as separate lines in this experiment. The Li-q (\SI{1580.0}{\electronvolt}: $[1s2p2s]\,^2P_{3/2} \rightarrow [1s^22s]\,^2S_{1/2}$) and Li-r (\SI{1579.6}{\electronvolt}: $[1s2p2s]\,^2P_{1/2} \rightarrow [1s^22s]\,^2S_{1/2}$) satellites also have similar energies and remain unresolved in \autoref{fig:Xray_spectra}.

The intensity and spectral position of the recorded lines exhibit modulations along the $z$-direction. This can be observed in \autoref{fig:Xray_spectra}, which shows a higher intensity of the lines for $z < \SI{0}{\milli \meter}$. In \autoref{fig:Xray_spectra}b, red crosses indicate the position of He-$\alpha$ IC line; the spectral position varies with $z$, and the magnitude of this deviation is $<\SI{1}{\electronvolt}$. We discuss potential reasons for the observed modulation in a later section. 

Although we only show data from MARZ1 and MARZ3 in \autoref{fig:Xray_spectra}, results from MARZ2 also exhibit the same emission lines, and the line ratios are consistent with that in MARZ1 and MARZ3. In {\color{blue} \S}\ref{sec:Xrayanalysis}, we use the line ratios of the observed He-like lines and Li-like satellites to constrain the density, temperature, and homogeneity of the emitting plasma in the reconnection layer.

\section{Discussion of Results}
\label{sec:discussion}

\subsection{Current and Magnetic Field Measurements}
\label{sec:magnetic_discussion}

In {\color{blue} \S}\ref{sec:current_measurements}, we observed equal current division of the MITL current between the two wire arrays. In addition, inductive probes measuring the advected magnetic field from separate arrays at the same radial distance (\SI{5}{\milli\meter} in MARZ1, \SI{20}{\milli\meter} in MARZ3) recorded similar magnetic field strength, consistent with equal current splitting (\autoref{fig:inductive_probes}). 

The advected magnetic field, however, does not reproduce the shape of the driving current, but instead exhibits a slower initial rise, followed by a faster ramping up later in time. This happens at around 320~ns for the 5~mm probes in MARZ1 and MARZ3, as seen in \autoref{fig:inductive_probes}(b and c). This effect was also observed in simulations of the experiment, and was found to be a consequence of a change in the wire ablation due to heating of the wire cores. \cite{datta2024simulations} In the simulations, the wire cores cool initially, but eventually begin to heat up due to re-absorption of emission generated by the surrounding plasma. The hotter cores are more conductive, and restrict the transport of magnetic field into the plasma flow from the cathode-wire gap. Although this effect is an important consequence of radiation transport observed both in the experiment and simulations \cite{datta2024simulations}, we note that the rise in the magnetic field occurs well after the onset of strong cooling in the experiments (around \SI{220}{\nano\second}), as discussed later in \S\ref{sec:cooling}.


\subsection{Density and temperature in the inflow region}
\label{sec:svs_snalaysis}

We estimate the ion density $n_i$ and electron temperature $T_e$ in the inflow to the reconnection layer by performing least-squares-fitting of synthetic spectra to the measured visible spectroscopy data shown in \autoref{fig:SVS}(a-d). The synthetic spectra $I_{\omega}$ are generated by solving the steady-state radiation transport equation (\autoref{eq:radtrans}) \cite{Drake2006} along the spectrometer’s line-of-sight (LOS) $s$ [see \autoref{fig:load_setup}], using spectral emissivity $\epsilon_\omega(n_i,T_e)$ and absorption opacity $\alpha_\omega(n_i,T_e)$ values calculated by PrismSPECT.\cite{macfarlane2013simulation} 
\begin{equation}
 \partial_s{I_\omega(s)} = \epsilon_\omega(s) - \alpha_\omega(s) I_\omega(s)
 \label{eq:radtrans}
\end{equation}
The PrismSPECT model uses a steady-state non-local thermodynamic equilibrium (nLTE) model with Maxwellian free electrons. The PrismSPECT calculations additionally assume a zero-width plasma with no background radiation field; comparison of finite width PrismSPECT simulation results with the zero-width results show that the effect of background radiation is negligible in this regime \citep{datta2023machine}.

As shown by the green line in \autoref{fig:load_setup}c, the diagnostic LOS samples plasma ablating from only one array in MARZ1. To solve radiation transport, we assume constant electron temperature along this LOS $s$. Previous experimental measurements in exploding wire arrays show little spatial variation in the temperature due to high thermal conductivity in pulsed-power-driven plasmas.\citep{russell2022perpendicular} Because the density falls with radial distance from the wires, \cite{burdiak2017structure,datta2022structure} we expect density along the LOS to peak at the center and fall towards the edges. The rocket model provides a simple description of the variation of mass density generated from wire arrays:\citep{lebedev2002snowplow,lebedev2004implosion}

\begin{equation}
    \rho\left(r, t^{\prime}\right)=\frac{\mu_0}{8 \pi^2 R_0 r V^2}\left[I\left(t^{\prime}-\frac{r-R_0}{V}\right)\right]^2
    \label{eq:rocket}
\end{equation}

Here, $r$ is the radial location around the wire array, $R_0 = \SI{20}{\milli \meter}$ is the radius of the wire array, and $V$ is the ablation velocity. A Gaussian function $n_i(s,t) = n_0(t) \exp[ (s-s_0)^2/ 2 \sigma(t)^2] $, with peak value $n_0(t)$ and standard deviation $\sigma(t)$ is a good approximation to the expected density along $s$ calculated from \autoref{eq:rocket}. Here, $s_0$ is the center of the diagnostic LOS. Furthermore, using the measured value of the flow velocity $V \approx\SI{110}{\kilo \meter \per \second}$ (see \autoref{fig:velocity}), we constrain the value of $\sigma(t)$ for our analysis, reducing the number of unknowns for fitting.

\begin{figure}[t!]
\includegraphics[page=6,width=0.5\textwidth]{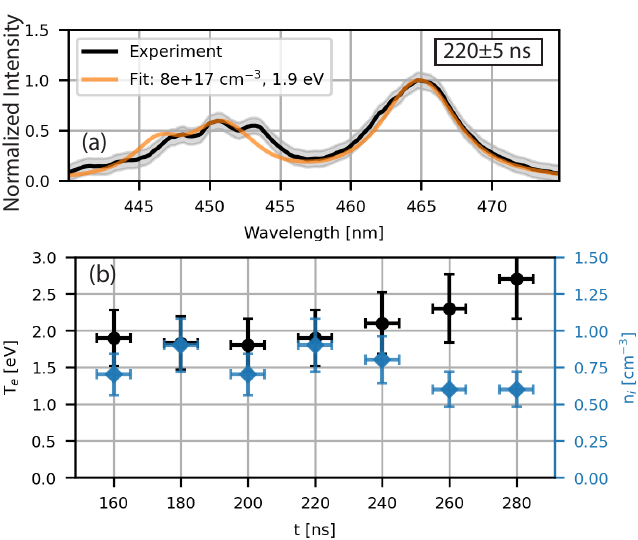}
\centering
\caption{\small (a) A least-squares fit of synthetic data (orange) generated from radiation transport and Prismspect simulations to the experimental spectrum (black) at $t=220\pm\SI{5}{\nano\second}$ for visible spectra collected at $\SI{8}{\milli \meter}$ from the wires. (b) Temporal variation of electron temperature (black) and ion density (blue) in the inflow region determined by fitting synthetic spectra to the visible spectroscopy data collected at $\SI{8}{\milli \meter}$ from the wires.
 }
\label{fig:SVS_analysis}
\end{figure}

\autoref{fig:SVS_analysis}a shows a synthetic fit (orange) to the measured spectrum (black) for the Al-II and Al-III inter-stage lines between 440-\SI{480}{\nano\meter} at $220 \pm \SI{5}{\nano \second}$, collected at $\SI{8}{\milli \meter}$ from the wires in MARZ1. The synthetic fit reproduces the experimental spectrum well. Ion density is sensitive to the width of the well-isolated Al-II \SI{466}{\nano \meter} line, and electron temperature is sensitive to the line ratio of the inter-stage Al-II \SI{466}{\nano \meter} and Al-III \SI{448}{\nano \meter} and \SI{452}{\nano \meter} lines. Temperature variations modify the relative population densities of the Al-II and Al-III ionization states. Increasing the temperature therefore increases the relative intensity of Al-III lines, while the Al-II lines become weaker, and completely disappear for temperatures $T_e > \SI{4}{\electronvolt}$, thus placing an upper bound on the electron temperature.

The time-resolved ion density and electron temperature determined from this analysis at $\SI{8}{\milli \meter}$ from the wires are shown in \autoref{fig:SVS_analysis}b. The electron temperature increases from about \SI{1.8}{\electronvolt} at \SI{160}{\nano\second} to \SI{2.6}{\electronvolt} later in time at \SI{280}{\nano\second}. The density remains roughly constant, varying between $5-\SI{8e17}{\per \centi \meter \cubed}$ between 160-280~ns. After this time, continuum emission dominates, and synthetic spectra indicate that the disappearance of the line emission is consistent with a rise in the ion density $n_i \gtrapprox \SI{3e18}{\per \centi \meter \cubed}$ and temperature $T_e \gtrapprox \SI{3}{\electronvolt}$. The increasing density is also consistent with the increasing total emission $\int I(\lambda) d\lambda$ measured at this time in \autoref{fig:SVS}a.  

Further from the array, at  \SI{17}{\milli \meter} from the wires (\autoref{fig:SVS}d), the electron temperature is found to be roughly \SI{2}{\electronvolt} between 220-\SI{380}{\nano \second}, and the ion density is lower, at about $2-\SI{6e17}{\per \centi \meter \cubed}$ further away from the wires. This shows that the temperature remains roughly constant in the plasma flows, while the density falls with radial distance from the wires, as expected due to divergence.

Finally, we perform synthetic spectral and radiation transport modeling to better understand the absorption features observed in MARZ3 [\autoref{fig:SVS}(e-f)], which samples plasma emission and absorption along a chord that includes both arrays and the layer plasma (blue line in \autoref{fig:load_setup}c). 
We model the plasma from each array with a Gaussian density (same peak value $n_0$ and variance $\sigma^2$) and a homogeneous temperature $T_0$. The layer is modeled as a region of thickness $\SI{1.5}{\milli\meter}$ with ion density $n_L$ and temperature $T_L$. A parametric study was performed by varying the layer density and temperature, and the synthetic modeling shows that the layer acts as a continuum backlighter, generating emission that is absorbed by the array plasma between the layer and the collection optics. 
The synthetic modeling allows us to qualitatively compare the layer density to the array density through inspection of emission and absorption features.
When the the layer density $n_L$ is comparable to the peak array plasma density $n_0$, the higher-opacity Al-II~624~nm line appears as an absorption feature, whereas the other Al lines appear as emission features, which is what is seen in the experiment. For layer densities much greater than the array plasma density, all of the Al lines appear as absorption features, whereas for layer densities less than the array density, all Al lines are emission features. Therefore, the presence of the Al-II~624~nm absorption feature in our experimental spectra  [(\autoref{fig:SVS}(e-f)] is consistent with the layer density being similar to the array plasma density at this location ($y = \SI{26.5}{\milli\meter}$), which is far away from the center of the layer. 
Spectra collected further downstream in MARZ3 ($y = \SI{35}{\milli\meter}$) only show emission features and no absorption features, indicating that the layer density decreases as the plasma flows away from the center of the reconnection layer, consistent with resistive MHD simulations of the experiment\cite{datta2024simulations}.


\subsection{Bow Shock Analysis}
\label{sec:bow_shock_discussion}
 
The measured Mach angle of the bow shock around the T-probe (\autoref{fig:bow_shock}) can provide information about the Mach number in the inflow region. The estimated Mach angle $\mu$ of the shock is about $\mu \approx \SI{30}{\degree}$, which corresponds to an upstream Mach number of about $M_{\text{up}} = 1/\sin(\mu) \approx 2$.  The shock stand-off distance, estimated from the width of the emission region at the leading edge of the probe, is about $\SI{0.2}{\milli\meter}$. The negligible change in the shock structure between 300-400~ns also indicates that the Mach number remains roughly constant. 

The resistive diffusion time of the magnetic field through the obstacle and the stagnated plasma is about $\tau_\eta \sim (\SI{1}{\milli\meter})^2/\bar{\eta}_{\text{glass}} + (\SI{0.2}{\milli\meter})^2/\bar{\eta}_{\text{plasma}} \sim 0.1-\SI{1}{\nano\second}$, which is smaller than the hydrodynamic time $L/V \approx \SI{5}{\nano\second}$. Here, we estimate the magnetic diffusivity $\bar{\eta}_{\text{plasma}}$ using Spitzer resistivity calculated with a temperature of $T_e\approx 2-\SI{10}{\electronvolt}$. The flow velocity at this time $V (t = \SI{350}{\nano\second}) \approx \SI{200}{\kilo \meter \per \second}$ is determined from the transit time of the plasma to the T-probe. Since the hydrodynamic time is comparable to the diffusion time, decoupling of the magnetic field and the plasma can result in hydrodynamic shock formation \citep{burdiak2017structure,datta2022structure}.  From the sonic Mach number $M_S = V / C_S \approx 2$ and the measured flow velocity, we estimate the ion sound speed $C_S \approx \sqrt{\bar{Z}T_e/m_i} \approx \SI{100}{\kilo\meter\per\second}$. However, this results in an estimated temperature of $\bar{Z}T_e \approx \SI{2}{\kilo \electronvolt}$, which is three orders of magnitude larger than the measured temperature in the inflow region from visible spectroscopy.

If, on the other hand, we assume that the shock is magnetohydrodynamic and magnetically dominated $\beta \ll 1$, then $M_A = V / V_A$, and the expected Alfvén speed is $V_A\approx B/\sqrt{\rho \mu_0} \approx \SI{100}{\kilo\meter\per\second}$. In {\color{blue} \S}\ref{sec:cooling_rate}, we show that the plasma $\beta \approx 0.1$ in the inflow region. From the measured value of the magnetic field ($B \approx 3$ T) at this time, the estimated ion density from the shock shape is $\SI{2e16}{\per \centi \meter \cubed}$, which is an order of magnitude lower than the expected density inferred from visible spectroscopy. We note that this is an upper bound on the density estimate from the shock shape, because the probe would measure a higher compressed magnetic field for MHD shock formation.

Therefore, the observed shock shape does not match the inflow conditions measured using inductive probes and visible spectroscopy. A potential cause of this discrepancy could be the generation of photo-ionized plasma from the probe surface, because of the harsh X-ray environment provided by the Z machine. Photo-ionized plasma at the T-probe tip may increase the post-shock pressure, creating a larger shock angle. Further investigation of this mismatch will require direct measurements of the post-shock density and temperature, and will be pursued in future experiments.

\subsection{Radiative Cooling and Generation of High-Energy Emission from the Reconnection Layer}
\label{sec:cooling}

\begin{figure}[b!]
\includegraphics[page=13,width=0.5\textwidth]{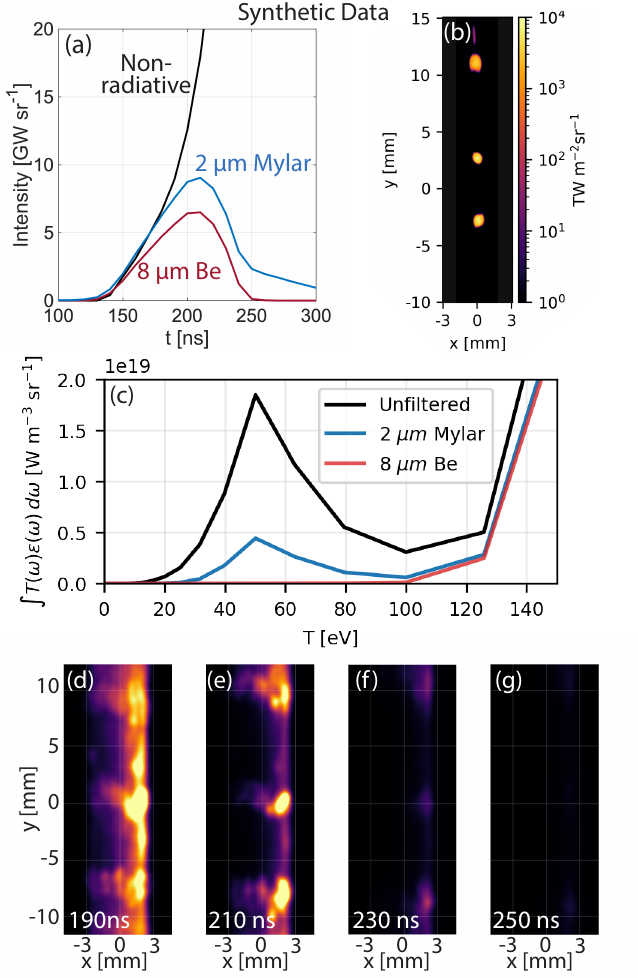}
\centering
\caption{\small Synthetic diagnostics from simulations of the MARZ experiment: (a) Filtered X-ray emission from the reconnection layer. The emission is filtered with \SI{8}{\micro\meter} Be (red) and \SI{2}{\micro\meter} Mylar (blue) to match the diode filters in the experiment. The black curve shows the expected X-ray emission (filtered with \SI{8}{\micro\meter} Be) for the simulation with no radiative cooling. (b) X-ray emission from the layer filtered with \SI{8}{\micro\meter} Be. Plasmoids primarily generate high-energy $>\SI{1}{\kilo\electronvolt}$ emission from the layer. (c) Filtered emissivity of the aluminum plasma generated using Spk tables. (d-g) Synthetic X-ray images of the reconnection layer, filtered with \SI{2}{\micro\meter} Mylar, and using the same line of sight as X-ray camera B in \autoref{fig:XrayImages}. X-ray emission from the layer decreases with time due to radiative cooling.
 }
\label{fig:Synthetic}
\end{figure}

The X-ray diode signals (\autoref{fig:diodes}) and the X-ray cameras (\autoref{fig:XrayImages}) both show a transient burst of high-energy X-ray emission from the reconnection layer. The initial rise in X-ray emission is consistent with increasing density and/or temperature of the reconnection layer during the formation stage. The temperature in the layer is initially high enough to generate high-energy X-rays with energies $>\SI{1}{keV}$. The subsequent fall in X-ray emission after \SI{220}{\nano\second} is consistent with rapid radiative cooling of the layer. The temporal change of X-ray intensity measured by the diodes also matches the intensity evolution observed in the X-ray images (\autoref{fig:XrayImages}).

We also observe a sharp fall in X-ray emission from the reconnection layer in radiative resistive MHD simulations of the MARZ experiment.\cite{datta2024simulations} In the simulations, radiative collapse of the current sheet, characterized by a sharp fall in the layer temperature and a simultaneous rise in density, begins around \SI{200}{\nano\second} after current start. In \autoref{fig:Synthetic}a, we plot synthetic diagnostic data, calculated from simulations, of the filtered X-ray emission generated from the reconnection layer as a function of time. The filtered X-ray emission is generated by post-processing the simulation results using radiation transport modeling in XP2 \citep{datta2024simulations}. We spatially integrate the output intensity, and filter it using transmission curves for both $\SI{8}{\micro\meter}$ Be and $\SI{2}{\micro\meter}$ Mylar, in order to match the diode filters in the experiment. In \autoref{fig:Synthetic}a, we additionally show the expected X-ray emission (filtered with $\SI{8}{\micro\meter}$ Be) for the case with no radiative cooling. 

In the absence of radiative cooling, X-ray emission from the layer would continue to rise, as the layer density ramps up in time at a consistently high temperature $>\SI{100}{\electronvolt}$. In the radiatively-cooled case, however, the emission peaks and then falls sharply, similar to that in the experiment. The temperature in the layer is initially $>\SI{100}{\electronvolt}$, which is high enough to generate $> \SI{1}{\kilo\electronvolt}$ X-rays. However, the subsequent fall in X-ray emission occurs due to radiative collapse of the layer, which rapidly cools as the density increases. The simulations therefore confirm that the rapid fall in X-ray emission is an experimental signature for strong radiative cooling of the reconnection layer. \autoref{fig:Synthetic}a also shows that the FWHM of the harder ($>\SI{1}{\kilo\electronvolt}$) emission is shorter than that of softer $>\SI{100}{\electronvolt}$ emission, similar to that in the experiment. This is expected, since as the temperature of the emitting plasma falls, the spectral distribution of the emitted radiation shifts to lower energies. Thus, the higher-energy emission falls earlier than the lower-energy emission.

Simultaneous measurements of X-ray emission by the \SI{8}{\micro\meter} Be and \SI{2}{\micro\meter} Mylar filtered diodes in the experiment (\autoref{fig:diodes}) provide constraints on the temperature of the emitting plasma. \autoref{fig:Synthetic}c shows the emissivity of an aluminum plasma with ion density \SI{5e18}{\per\centi\meter\cubed} as a function of the plasma temperature, calculated using the atomic code Spk.\cite{crilly2020simulation} Spk uses a nLTE model and includes line, recombination, and bremsstrahlung emission.\cite{datta2024simulations} The unfiltered emissivity exhibits a smaller peak around $T \approx \SI{50}{\electronvolt}$; the emissivity is lower between 50-100~eV, and increases for $T > \SI{100}{\electronvolt}$. The smaller peak results from L-shell line emission of photons with energies of $100-\SI{300}{\electronvolt}$, while the increased emission at temperatures $T >\SI{100}{\electronvolt}$ is due to higher energy $>\SI{1}{\kilo \electronvolt}$ Al K-shell emission. We also show the emissivity filtered using X-ray transmission tables for \SI{8}{\micro\meter} Be and \SI{2}{\micro\meter} Mylar in \autoref{fig:Synthetic}c. The \SI{8}{\micro\meter} Be filter significantly attenuates radiation with energies $<\SI{1}{\kilo \electronvolt}$, whereas the \SI{2}{\micro\meter} Mylar filter exhibits a smaller window of transmission around photon energies $\approx \SI{200}{\electronvolt}$. Therefore, the filtered emissivity through \SI{2}{\micro\meter} Mylar is significant for temperatures $T > \SI{25}{\electronvolt}$, while for \SI{8}{\micro\meter} Be the signal is only significant for temperatures $T > \SI{100}{\electronvolt}$.\footnote{We also solve the 1D radiation transport equation (\autoref{eq:radtrans}) to determine the filtered intensity of radiation from the emitting plasma, and the conclusions presented here are consistent with that seen from the emissivity in \autoref{fig:Synthetic}c.}

The different responses of the filtered diodes shown in \autoref{fig:Synthetic}c allows us to constrain the temperature of the emitting plasma. Initially, in region A (see \autoref{fig:diodes}), where neither diode records any signal, we expect the plasma temperature to be $T < \SI{25}{\electronvolt}$. In region B, where the Mylar diode records signal, while the Be diode does not, the temperature of the emitting plasma is constrained to be $25 < T < \SI{100}{\electronvolt}$. In C, where both diodes simultaneously record signals, the temperature must be $T > \SI{100}{\electronvolt}$. Similarly, the expected temperature is $25 < T < \SI{100}{\electronvolt}$, and $T < \SI{25}{\electronvolt}$ in regions D and E respectively. Therefore, the diode signals indicate an initial heating of the reconnection layer to temperatures $> \SI{100}{\electronvolt}$, followed by a sharp fall below $T < \SI{25}{\electronvolt}$ due to radiative cooling. We show later that the expected temperature at the time of peak emission (region C), is consistent with results from X-ray spectroscopy analysis in {\color{blue} \S}\ref{sec:Xrayanalysis}.

The diodes collect time-resolved emission integrated over the entire reconnection layer. 
The time-gated X-ray images (\autoref{fig:XrayImages}) complement this measurement by providing spatial resolution at discrete times. 
These images were also filtered with \SI{2}{\micro\meter} Mylar, and at around 190-200~ns both the layer and hotspots generate emission in the spectral range transmitted by the Mylar filter, indicating both the hotspots and the layer have $T > \SI{25}{\electronvolt}$.
Later in time (230-240~ns), the hotspots remain bright, while emission from the layer has significantly decreased. This indicates that the layer has cooled to $T < \SI{25}{\electronvolt}$, while the hotspots have remained above $\SI{25}{\electronvolt}$, such that the signal on the Mylar filtered diode is dominated by hotspot emission. 
By \SI{250}{\nano\second}, there is no emission recorded on the X-ray camera or the Mylar filtered diode, consistent with both the hotspots and the layer cooling to $T<\SI{25}{\electronvolt}$.

\autoref{fig:Synthetic}(d-g) show synthetic X-ray images of the reconnection layer generated from post-processing the 3D simulation results in XP2 between 190-\SI{250}{\nano\second}. The synthetic images are filtered with $\SI{2}{\micro\meter}$ Mylar, and use the same LOS as X-ray camera B in the experiment (see \autoref{fig:load_setup}d). As observed in \autoref{fig:Synthetic}(d-g), emission from the reconnection layer falls as a result of radiative cooling, consistent with the experimental images in \autoref{fig:XrayImages}. The synthetic images also show hotspots of emission within the elongated reconnection layer, similar to that in the experiment. These hotspots correspond to the position of magnetic islands or plasmoids in the simulations, which are generated by the tearing instability.\citep{loureiro2007instability,uzdensky2010fast} The plasmoids appear as localized regions of enhanced emission due to their relatively higher electron density and temperature.\cite{datta2024simulations} Enhanced emission from plasmoids has also been reported in relativistic-PIC simulations of extreme astrophysical objects.\citep{schoeffler2019bright} 

The majority of the high-energy emission is generated by the plasmoids in our simulations because of their higher temperature. This can be observed in \autoref{fig:Synthetic}b, which shows X-ray emission from the reconnection layer filtered with $\SI{8}{\micro\meter}$ Be. Emission recorded by the \SI{8}{\micro\meter} Be filtered diode in the experiment may thus be dominated by higher energy emission from the hotspots with temperature $T > \SI{100}{\electronvolt}$. This is further supported by the analysis of X-ray spectroscopy from the experiment, which is provided in the following subsection. 

\begin{figure}[t!]
\includegraphics[page=12,width=0.5\textwidth]{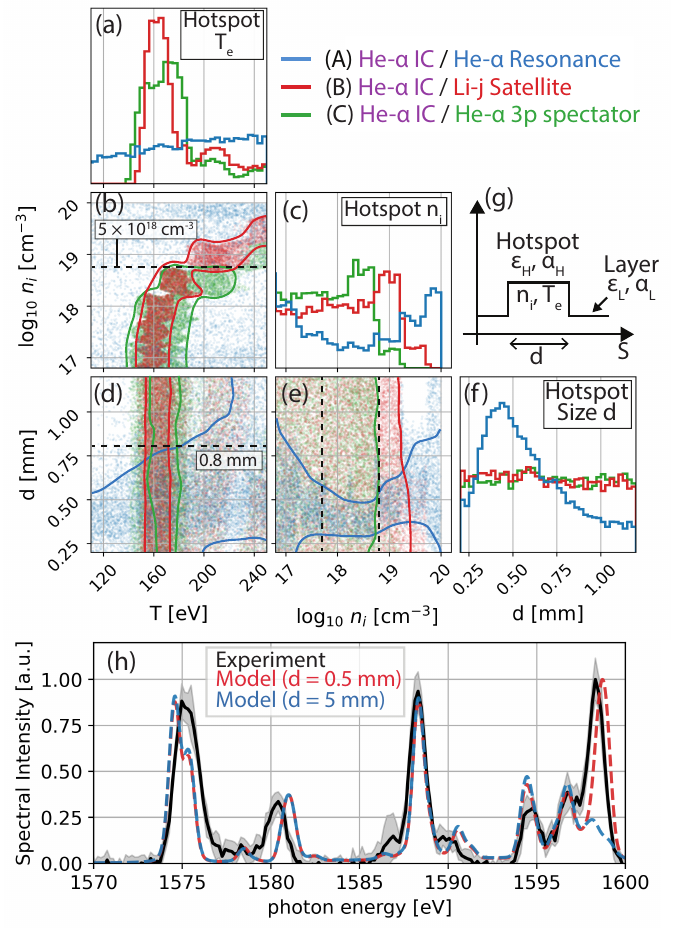}
\centering
\caption{\small Corner plot of solutions which match  the line ratios of (A) He-$\alpha$ IC / He-$\alpha$ Resonance (blue), (B) He-$\alpha$ IC / Li-j Satellite (red), and (C) He-$\alpha$ IC / He-$\alpha$ with 3p spectator (green). (b, d, and e) show the 2D scatter plots for $n_i$, $T_e$, and $d$; (a, c, and f) show the probability density distributions of these values for valid solutions. Solid lines represent contours enclosing 90\% of the solutions. (g) Radiation transport model for emission from a single hotspot with emissivity $\epsilon_H$ and opacity $\alpha_H$, embedded in a homogeneous layer with $\epsilon_L$ and opacity $\alpha_L$. (h) Comparison of synthetic spectra with the experimental spectrum ($z=\SI{10}{\milli\meter}$, MARZ3). Both synthetic spectra are calculated for $T = \SI{170}{\electronvolt}$ and $n_i = \SI{1e18}{\per \centi \meter \cubed}$, but using sizes of $d = \SI{0.5}{\milli \meter}$ (red) and $d = \SI{5}{\milli \meter}$ (blue) respectively. }
\label{fig:MC}
\end{figure}

\subsection{Analysis of X-Ray Spectroscopy}
\label{sec:Xrayanalysis}

We estimate the temperature and density of the plasma in the reconnection layer by comparing the measured spectra shown in \autoref{fig:Xray_spectra} with synthetic spectra. We use the atomic code SCRAM to generate nLTE spectral emissivity $\epsilon_\omega(n_i,T_e)$ and absorption opacity $\alpha_\omega(n_i,T_e)$ tables\cite{hansen2007hybrid}, which are then used to solve the radiation transport equation \cite{Drake2006} (\autoref{eq:radtrans}) along the diagnostic LOS to model the output intensity spectrum. For this analysis, SCRAM includes spectral line broadening effects (Stark and thermal Doppler), and incorporates a background radiation field by assuming a homogeneous cylindrical plasma of diameter $\SI{1}{\milli \meter}$. A characteristic length of \SI{1}{\milli\meter} is chosen because it is comparable to the width of the X-ray emission region, as observed in X-ray images of the reconnection layer (\autoref{fig:XrayImages}, \autoref{fig:XrayImages}c). The contribution of scattering to the total opacity is expected to be small in this regime, and is thus, excluded from the SCRAM calculations.

The SCRAM results show that Al K-shell lines do not appear for temperatures below \SI{60}{\electronvolt}. Furthermore, as expected for inter-stage transitions,\cite{sobelman2012atomic} the relative emissivities of the Li-like satellites and He-like lines are strong functions of temperature. Li-like satellites exhibit higher emissivities at lower temperatures, whereas the emissivity of He-like lines dominates at higher temperatures. The He-$\alpha$ resonance transition exhibits much higher emissivities and opacities compared to the other He-like lines, consistent with the relatively higher probability of the resonance transition \citep{herzberg1944atomic,sobelman2012atomic}. At $T_e \approx \SI{100}{\electronvolt}$, the Li-j,k,l satellites and the  He-$\alpha$ resonance line have similar emissivity values; however, the opacity of the He-$\alpha$ resonance line is several orders of magnitude higher than both the Li-like satellites and other He-like lines. For instance, at $n_i = \SI{1e18}{\per\centi\meter\cubed}$ and $T_e=\SI{100}{\electronvolt}$, the attenuation length scale $\alpha_\omega^{-1}$ for the He-$\alpha$ resonance line is about \SI{0.1}{\milli\meter}, while that for the other He-like lines and Li-like satellites is $>\SI{10}{\milli\meter}$. Therefore, due to the high opacity of the He-$\alpha$ resonance line, radiation transport calculations are required to accurately model the spectral intensity of the emission from the reconnection layer.

To solve the 1D radiation transport problem along the diagnostic LOS, we first assume emission from a plasma of length $d$ with homogeneous ion density $n_i$ and electron temperature $T$. Values of $n_i$, $T$, and $d$ are then randomly sampled from uniform distributions to find solutions that match within $20\%$ the experimentally observed line ratios. The model includes the effect of source and instrument broadening, but neglects Doppler shift, which is $<0.25 \, \text{eV}$, as calculated from the hotspot velocities in \autoref{fig:UXI_velocity}. In particular, we compare three different line ratios shown in \autoref{fig:Xray_spectra}d --- (A) He-$\alpha$ IC / He-$\alpha$ Resonance (blue), (B) He-$\alpha$ IC / Li-j Satellite (red), and (C) He-$\alpha$ IC / He-$\alpha$ with 3p spectator (green). \autoref{fig:MC} shows a corner plot of the solutions that individually satisfy these line ratios at $z = \SI{10}{\milli\meter}$ in MARZ3 (see \autoref{fig:Xray_spectra}d). The corner plot shows one and two-dimensional projections of the three-dimensional parameter space ----\autoref{fig:MC}(b, d, and e) are 2D scatter plots for solutions, while \autoref{fig:MC}(a,c and f) show the probability distribution functions for the values of $n_i$, $T$ and $d$ of the solutions. The solid contours in \autoref{fig:MC}(b, d, and e) enclose 90\% of the solutions. The intersection of the solutions for these line ratios constrains the properties of the emitting plasma.

As seen in \autoref{fig:MC}b, which shows a scatter plot of the ion density against temperature, solutions for the B (red) and C (green) line ratios constrain $T_e$ to a narrow band around $T \approx 170\pm\SI{20}{\electronvolt}$, and provide an upper bound of about $n_i \lesssim \SI{5e18}{\per\centi\meter\cubed}$ on the ion density (indicated via the black dashed line). These bounds are determined from the intersection of these solutions. The size of the plasma $d$ is poorly constrained from these line ratios, since line ratios B and C are for the optically thin lines, and optical depth $(\alpha_{\omega}(n_i,T_e) d)$ thus has a limited effect. \autoref{fig:MC}(d and e) also clearly show overlapping solutions at $T \approx 170\pm\SI{20}{\electronvolt}$ and $n_i \lesssim \SI{5e18}{\per\centi\meter\cubed}$ respectively, but exhibit a wide range of possible values for $d$. 

The size of the emitting plasma can, however, be constrained through solutions for line ratio A (blue). Line ratio A is the relative intensity of the He-$\alpha$ IC compared to the higher opacity He-$\alpha$ resonance line, which is a strong function of optical depth. This line ratio is well satisfied for a wide range of $T$, $n_i$ and $d$; however, from \autoref{fig:MC}d, we observe that for temperatures $T = 170\pm \SI{20}{\electronvolt}$ that satisfy line ratios B and C, line ratio A is only satisfied for $d \leq \SI{0.8}{\milli\meter}$ (indicated via the black dashed line). If the size of the emitting plasma exceeds this value, we can no longer satisfy all line ratios simultaneously because of over-damping of the high-opacity He-$\alpha$ resonance line. 
Notably this value for $d$ is significantly lower than the length of the reconnection layer observed in \autoref{fig:XrayImages}c, which implies that only a small region of the layer is contributing to the the emissivity and opacity of the system for these photon energies.

This is further illustrated in \autoref{fig:MC}h, which compares two synthetic spectra to the experimental spectrum at $z=\SI{10}{\milli\meter}$ (MARZ3). Both synthetic spectra are calculated for $T = \SI{170}{\electronvolt}$ and $n_i = \SI{1e18}{\per \centi \meter \cubed}$, but using sizes of $d = \SI{0.5}{\milli \meter}$ (red) and $d = \SI{5}{\milli \meter}$ (blue) respectively. As expected, the synthetic spectrum for $d = \SI{0.5}{\milli \meter}$ reproduces the line ratios and line widths of the experimental spectrum well.  The synthetic spectrum for $d = \SI{5}{\milli \meter}$ reproduces the relative intensities of the Li-like satellites and the He-$\alpha$ IC and spectator transitions, but fails to reproduce the He-$\alpha$ resonance line. This analysis therefore indicates that Al K-shell emission from the reconnection layer is predominantly generated by sub-millimeter size hotspots in the layer. This is further supported by the X-ray images of the layer (\autoref{fig:XrayImages}) that show the presence of strongly-emitting hotspots of size $<\SI{1}{\milli\meter}$, as well as by the simulations, which show strong localized emission of $>\SI{1}{\kilo\electronvolt}$ photons from the plasmoids (\autoref{fig:Synthetic}b). Assuming that the hotspot density lies between the upper bound of $n_i \leq \SI{5e18}{\per \centi \meter \cubed}$ and the lower bound of $n_i \geq \SI{5e17}{\per \centi \meter \cubed}$, which is the inflow density from visible spectroscopy (\autoref{fig:SVS_analysis}), we find, from \autoref{fig:MC}e, that $0.3 \leq d \leq \SI{0.5}{\milli \meter}$, consistent with the plasmoid widths observed in simulations.\cite{datta2024simulations}

To estimate an upper bound on the bulk temperature of the layer, we consider a simple model that includes a single hotspot of size $d$ embedded in a homogeneous layer with emissivity $\epsilon_L$ and opacity $\alpha_L$, as shown in \autoref{fig:MC}g. Radiation transport solutions for this model indicate that the bulk layer temperature must be less than about $T_{\text{bulk}} \leq \SI{75}{\electronvolt}$. At around $T_{\text{bulk}} \approx \SI{75}{\electronvolt}$, contribution from the layer is not negligible and modifies the spectral intensity from the hotspots; however, solutions that satisfy the line ratios in the experiment are still possible. At higher layer temperatures, the He-$\alpha$ resonance line becomes strongly absorbed by the layer, and the experimental spectrum can no longer be reproduced. 

\autoref{fig:Xray_spectra} shows that the hotspots form elongated structures of length $\sim \SI{10}{\milli\meter}$ along $z$. The above analysis was repeated at multiple $z$-positions for the different shots. Although bounds on the inferred hotspot density, temperature, and size show slight variations along $z$, the values remain largely consistent. This is unsurprising because the same Al K-shell lines are observed at different $z$ and the line ratios remain largely similar, despite small modulations, as seen in \autoref{fig:Xray_spectra}. The modulation in the spectral position of the lines is unlikely to be caused by Doppler shift --- as mentioned before, the maximum Doppler shift expected is \SI{0.25}{\electronvolt}, which is smaller than the \SI{1}{\electronvolt} modulation seen in \autoref{fig:Xray_spectra}. The spectral modulations could be a result of modulations in the position of the hotspots in the object plane along $x$. Deviations in the source position can lead to deviations in the position of the lines recorded on the image plate. From ray tracing simulations,\cite{harding2015analysis} the observed deviation in spectral position corresponds to a roughly \SI{1}{\milli\meter} deviation in the position of the hotspots. This deviation is comparable to the amplitude of the modulations in the plasmoid position generated by the MHD kink instability in the 3D simulation of the experiment.\cite{datta2024simulations} Thus, the spectral deviation of the lines along $z$ may be a preliminary indication of the MHD kink instability of the plasmoids. 

\subsection{Cooling Rate}
\label{sec:cooling_rate}

Analysis of experimental data in the previous sections has provided quantitative estimates of the plasma properties. In this section, we use these experimental results to further characterize the net cooling rate in the reconnection layer at the time of the onset of strong radiative cooling.

The temporal evolution of the reconnection layer temperature depends on the relative magnitudes of the terms in the internal energy equation:\citep{goedbloed_keppens_poedts_2010}

\begin{equation}
    \frac{\partial}{\partial t}(\rho \epsilon)+\nabla \cdot(\rho \epsilon \mathbf{v}) = \eta|j|^2 - p \nabla \cdot \mathbf{v} + \nabla {\bf v} : {\bf \tau} - P_{rad}
\label{eq:energy_eqn}
\end{equation}

\begin{table}\centering
\ra{1.3}
\caption{\small Estimated magnitudes of terms in the energy equation for the reconnection layer}
\begin{tabular}{|c|c|c|}
\hline
Term & Estimate & Value [\SI{}{\watt\per\meter\cubed}] \\
\hline
$\eta |j|^2$ & $\eta \left(B_{in} / \mu_0 \delta_{SP}\right)^2$ & $\SI{2e15}{}$ \\
$-p\nabla \cdot {\bf v}$ & $p_L(V_{in}/\delta_{SP} - V_{out}/L)$ & $\SI{6e15}{}$ \\
${\bf \nabla v : \tau}$ & $\mu(V_{in}/\delta_{SP})^2$ & $\SI{1e9}{}$ \\
$\nabla \cdot (\rho \epsilon {\bf v})$ & $(p_{in}V_{in}/\delta_{SP}-p_L V_{out}/L) /(\gamma -1)$ & $\SI{3e15}{}$ \\
$P_{rad}$ & \autoref{eq:isotropic_plasma} & $\SI{1e18}{}$ \\

\hline
\end{tabular}
\label{tab:energy_eqn}
\end{table}

Here, $\rho \epsilon = p /(\gamma-1)$ is the internal energy density,  $\nabla \cdot(\rho \epsilon \mathbf{v})$ is the advective term, $\eta|j|^2$, $- p \nabla \cdot \mathbf{v}$, and $\nabla {\bf v} : {\bf \tau}$ are the compressional, resistive, and viscous heating terms respectively. Lastly, $P_{rad}$ is the volumetric radiative loss from the layer. We can estimate the order-of-magnitude of these terms based on the plasma parameters in the layer and in the inflow, as shown in the second column of \autoref{tab:energy_eqn}. Here, $B_{in}, \, p_{in}$ and $V_{in}$ are the magnetic field, thermal pressure, and velocity in the inflow just outside the layer, and $p_L$ and $V_{out}$ are the layer pressure and the reconnection outflow velocity. The half-length and the half-width of the layer are denoted by $L$ and $\delta$ respectively. Lastly, $\eta$ and $\mu$ are the (Spitzer) resistivity and plasma viscosity \cite{richardsonnrl2019}, while $\gamma$ is the adiabatic index\citep{Drake2006}. 

To determine the net cooling rate, we require estimates of the inflow and layer parameters listed above. In our previous paper, Ref.~\onlinecite{datta2024plasmoid}, we presented approximations of these parameters at the time of onset of radiative cooling ($t \approx \SI{220}{\nano\second}$). We do not repeat this analysis here, but instead, summarize the key results. Based on the measured values of ion density $n_i$, electron temperature $T_e$ (see {\color{blue} \S}\ref{sec:svs_snalaysis}), magnetic field $B$, and flow velocity $V$ (see {\color{blue} \S}\ref{sec:probes}) in the inflow region, we find that the inflows are super-Alfvénic ($M_A = V / V_A \approx 7$). We reiterate that these quantities are measured on the side of the array opposite the reconnection layer. Therefore, by taking advantage of the azimuthal symmetry of the outflows from the arrays (which form the inflows to the layer), we diagnose the inflow conditions, unaffected by the reconnection process. Super-Alfvénic inflows have previously been shown to generate shock-mediated magnetic flux pile-up upstream of the reconnection layer, dividing the inflow into pre-shock and post-shock pile-up regions. \cite{suttle2016structure,suttle2018ion,fox2011fast,olson2021regulation} In simulations of the MARZ experiment, magnetic flux pile-up results in fast perpendicular MHD shocks upstream of the reconnection layer.\citep{datta2024simulations} \, Ref.~\onlinecite{datta2024plasmoid} estimates the parameters in the post-shock inflow region by solving the Rankine-Hugoniot jump conditions, the solution to which is a function of the upstream $M_A$, plasma beta $\beta$, and adiabatic index $\gamma$. \citep{goedbloed_keppens_poedts_2010} The plasma parameters in the pre-shock and post-shock inflow regions are listed in the first and second columns of \autoref{tab:table}.

The plasma properties in the post-shock pile-up region set the inflow conditions just outside the reconnection layer. Ref.~\onlinecite{datta2024plasmoid} uses two additional assumptions, both supported by simulation results, to approximate properties in the reconnection layer --- (1) the reconnection layer exists in pressure balance with the post pile-up inflow region; and (2) at \SI{220}{\nano\second}, which is right at the onset of strong cooling, there is negligible compression of the layer. The first assumption simply states that the kinetic $\rho_{in} V_{in}^2 / 2$, magnetic $B_{in}^2/2\mu_0$ and thermal $p_{in}$ pressures  just outside the reconnection layer must be balanced by the thermal pressure $p_L$ inside the layer. The second assumption argues that the temperature of the layer $T_L$ has not fallen enough at $\SI{220}{\nano\second}$ to trigger the radiative collapse and run-away density compression of the reconnection layer. The estimated plasma properties in the reconnection layer are listed in the third column of \autoref{tab:table}. We note that the estimated layer temperature at this time is $T_{bulk} \approx \SI{60}{\electronvolt}$, which is consistent with the upper bound ($T_{e,bulk} \lesssim \SI{75}{\electronvolt}$) determined from X-ray spectroscopy (\S\ref{sec:Xrayanalysis}), and the predicted ion density is $n_i \approx \SI{6e18}{\per \centi \meter \cubed}$, slightly greater than the upper bound of $n_i \lesssim  \SI{5e18}{\per \centi \meter \cubed}$) from X-ray spectroscopy. The estimated Lundquist number is $S_L = L V_A/\bar{\eta}\approx 120$, and the predicted Sweet-Parker layer half-width is $\delta_{SP} \approx L (S_L)^{1/2} \approx \SI{1.4}{\milli \meter}$ \cite{parker1957sweet,yamada2010magnetic}, which is comparable to the FWHM$=1.6\pm\SI{0.5}{\milli\meter}$ of the X-ray emission region observed in the time-integrated image of the reconnection layer (\autoref{fig:XrayImages}c). We approximate the layer half-length $L = \SI{15}{\milli \meter}$, using half the radius of curvature of the magnetic field lines at the mid-plane. Lastly, an estimate of the reconnection rate is determined by comparing the post-shock inflow velocity $V_{in} \approx \SI{20}{\kilo \meter \per \second}$ to the outflow velocity $V_{out} \approx \SI{72}{\kilo\meter\per \second}$,  estimated by extrapolating the linear trend in the hotspot velocity observed in \autoref{fig:UXI_velocity}. The inferred reconnection rate at this time is $V_{in}/V_{out} \approx 0.3$. This is roughly comparable to the expected reconnection rate from Sweet-Parker theory $S_L^{-1/2} \approx 0.1$ \cite{parker1957sweet}.

\begin{table}\centering
\ra{1.3}
\caption{\small Plasma parameters at the time of onset of radiative cooling (220~ns). Bold values are measured experimentally, while others are estimated/inferred. In column 1, we report values of $n_i$,  $T_e$, and $\bar{Z}$ from visible spectroscopy analysis at \SI{8}{\milli\meter} from the wires (see {\color{blue} \S}\ref{sec:svs_snalaysis}), magnetic field from averaging the values recorded by the inductive probes at $\SI{5}{\milli\meter}$ and $\SI{10}{\milli\meter}$ in MARZ3 (\autoref{fig:inductive_probes}c), and the flow velocity is estimated from the transit time of the magnetic field between the two probes (\autoref{fig:velocity}d). Parenthetical values show bounds from X-ray spectroscopy (\S \ref{sec:Xrayanalysis})}
\begin{tabular}{p{2.3cm}ccc}
\hline
\multicolumn{1}{c}{\textbf{Parameter}} & \textbf{Pre-shock} & \textbf{Post-shock} & \textbf{Reconnection} \\
\multicolumn{1}{c}{\textbf{}} & \textbf{Inflow} & \textbf{Inflow} & \textbf{Layer} \\
\hline
$n_i \, [\times \SI{e18}{\per \centi \meter \cubed}$] & {\bf 0.8} & 6 & 6 $(\mathbf{\lesssim 5})$ \\
$T_e$ [eV] & {\bf 1.9} & 30 & 60 $(\mathbf{\lesssim 75})$ \\
$\bar{Z}$ & {\bf 2} & 8 & 10 \\
$B_y$ [T] & {\bf 3.9} & 30 & - \\
$p$ [MPa] & 0.6 & 300 & 700 \\
$V_x$ [km/s] & {\bf 140} & 20 & - \\
$V_y$ [km/s] & - & - & {\bf 72} \\
$V_A$ [km/s] & 20 & 50 & - \\
$C_S$ [km/s] & 5 & 30 & 50 \\
$\beta$ & 0.1 & 0.8 & - \\
$\beta_{\text{kin}}$ & 60 & 0.1 & - \\
$\gamma$  & 1.2 & 1.1 & 1.1 \\
$d_i$ [mm] & 0.7 & 0.1 & 0.1 \\
$\lambda_{ii}$ [nm] & 20 & 1 & 1 \\
$S_L$ & - & - & 120 \\
$\tau_{E}$ [ns] & 2 & 1 & 1 \\
\hline
\end{tabular}
\label{tab:table}
\end{table}

Using the quantities listed in \autoref{tab:table}, we now estimate the relative magnitudes of the terms in the energy equation (\autoref{eq:energy_eqn}). As observed in the third column of \autoref{tab:energy_eqn}, Ohmic heating, compressional heating, and net enthalpy advection into the layer, are estimated to have similar magnitudes, whereas the contribution of viscous heating is small. We also note that comparing the conductive heat flux $-\kappa \nabla T \sim \kappa (T_L - T_{in}) / \delta_{SP}$ with the advective flux $p_{in}V_{in}/(\gamma-1)$ along the $x-$direction shows that conduction losses from the layer to the upstream inflow are expected to be small ($q_{cond,x}/q_{adv,x} < 0.1$). Here, $\kappa$ is the parallel thermal conductivity in the layer, which is a function of the layer density and temperature $T_L$, \cite{richardsonnrl2019}  and $T_{in}$ is the inflow temperature.

We estimate the radiative loss rate $P_{rad}$ from the layer by solving the radiation transport equation (\autoref{eq:radtrans}) for an isotropically emitting and absorbing medium.\citep{datta2024simulations,crilly2020simulation}

\begin{equation}
     P_{\text{rad}} = \frac{3}{4R} \int \frac{4\pi \epsilon_\omega}{\alpha_\omega}\left[1+\frac{2}{\tau_\omega^2}\left\{(1+\tau_\omega)e^{-\tau_\omega}-1\right\}\right] d\omega 
     \label{eq:isotropic_plasma}
\end{equation}

Here, $\epsilon_\omega$ and $\alpha_\omega$ are the spectral emissivities and opacities respectively, and $\tau_\omega = 2\alpha_\omega R$ is the optical depth. $R$ is the characteristic distance traveled by the radiation leaving the reconnection layer, which we approximate using the volume-to-surface area ratio $R = (1/\delta_{SP} + 2/L)^{-1}$ of a cuboidal slab of width $2\delta_{SP}$, and height and length $2L$. Spectral emissivity and opacity values for the estimated layer temperature and density are determined from SpK \cite{crilly2022spk}. The radiative loss rate calculated using \autoref{eq:isotropic_plasma} is $P_{\text{rad}} \sim \SI{e18}{\watt \per \meter \cubed}$, which is significantly larger than the total heating rate $P_{heat} \sim \SI{e16}{\watt \per \meter \cubed}$. The dimensionless cooling parameter is $R_{\text{cool}} \equiv \tau_{\text{cool}}^{-1} / \tau_A^{-1} \approx 400$ at \SI{220}{\nano\second}. Here, $\tau_{\text{cool}} = p_L/(P_{rad}-P_{heat})$ and $\tau_{A} = L / V_{A,in}$ are the net cooling and Alfvén transit times respectively. Radiative cooling therefore dominates heating in the experiment, which is consistent with the strong cooling inferred from the X-ray diagnostics of the reconnection layer (see {\color{blue} \S}\ref{sec:cooling}).

\section{Conclusions}
\label{sec:conclusions}
We present results from a strongly-radiatively cooled pulsed-power-driven magnetic reconnection experiment. Unlike previous pulsed-power-driven experiments, which either achieved strong cooling without plasmoid formation at low Lundquist numbers ($S_L < 10$),\cite{suttle2016structure,suttle2018ion} or insignificant cooling with plasmoid formation at higher Lundquist numbers ($S_L \sim 100$),\cite{hare2017formation,hare2018experimental} here we simultaneously achieve strong cooling $R \gg 1$ and plasmoid formation at $S_L > 100$.  Using a suite of current, inflow, and reconnection layer diagnostics, as described in {\color{blue} \S}\ref{sec:diagnostic_setup}, we obtain the following key results: 
 (1)~X-ray emission from the reconnection layer, as measured by the X-ray diodes and time-gated X-ray cameras, rises initially and then falls sharply, consistent with rapid cooling of the layer (see {\color{blue} \S}\ref{sec:cooling}). A similar effect is seen in radiative resistive magnetohydrodynamic simulations of the experiment, which show that X-ray emission would simply continue to rise without radiative cooling.\cite{datta2024simulations} (2)~The reconnection layer exhibits sub-millimeter size localized regions of strong X-ray emission (see {\color{blue} \S}\ref{sec:uxi_results}). These hotspots are consistent with the presence of plasmoids generated by the tearing instability in our simulations,\cite{datta2024simulations} which generate strong X-ray emission due to their higher electron density and temperature compared to the rest of the layer. (3)~These hotspots generate the majority of the high-energy X-ray emission from the layer; this is supported by X-ray images of the layer, and by X-ray spectroscopy, which shows that the measured X-ray spectra can be best explained by emitting regions of size $<\SI{1}{\milli\meter}$ (see {\color{blue} \S}\ref{sec:Xrayanalysis}). The generation of high-energy emission predominantly from the plasmoids is also observed in the simulations, as shown in \autoref{fig:Synthetic}.

Using the measured quantities, we infer key physical and dimensionless parameters describing the plasma in the inflow region and the reconnection layer in {\color{blue} \S}\ref{sec:cooling_rate}.  The estimated values of layer density, temperature, and size are consistent with bounds determined from experimental diagnostics. Finally, using these inferred quantities, we compare the various terms in the energy equation (\autoref{tab:energy_eqn}), and show that radiative loss is expected to dominate the total heating rate inside the reconnection layer ($R_{\text{cool}} > 100$). This further supports the strong cooling observed in the experiment.

Strong radiative cooling is expected to result in radiative collapse of the current sheet, similar to that in the simulations.\cite{datta2023plasma} In future experiments, we will further characterize the collapse of the reconnection layer at later times using time-resolved measurements of the density, size, and compression of the layer. Our current experiments additionally provide preliminary evidence of the 3D modulations in the structure of the plasmoids, and future experiments will further probe this 3D structure using time-resolved laser imaging. 

The results in this paper provide insight into a regime of magnetic reconnection that has not been previously explored in laboratory experiments. Our detailed measurements of the entire reconnection process provide high-quality data for benchmarking atomic codes and radiative-magnetohydrodynamic simulations. Our findings on the generation of transient X-rays bursts from the layer, and the localization of the high-energy emission within the plasmoids, are relevant to understanding observations of reconnection in remote and extreme astrophysical environments. Finally, the MARZ platform exhibits rich radiative physics, allowing for the investigation of radiation transport and cooling effects in HED plasmas beyond their relevance to magnetic reconnection.

\section{Acknowledgements}
The authors would like to thank the Z machine operations teams and the target fabrication team for their contributions to this work. Experimental time on the Z facility was provided through the Z Fundamental Science Program.
This work was funded by NSF and NNSA under grant no. PHY2108050, by the EAGER grant no. PHY2213898. RD acknowledges support from the MIT MathWorks and the MIT College of Engineering Exponent fellowships. DAU gratefully acknowledges support from NASA grants 80NSSC20K0545 and 80NSSC22K0828.
This work was supported by Sandia National Laboratories, a multimission laboratory managed and operated by the National Technology and Engineering Solutions of Sandia, LLC, a wholly owned subsidiary of Honeywell International, Inc., for the U.S. Department of Energy’s National Nuclear Security Administration under Contract No. DE-NA0003525.
This paper describes objective technical results and analysis. Any subjective views or opinions that might be expressed in the paper do not necessarily represent the views of the U.S. Department of Energy or the United States Government. 

\section{Declaration of Conflicts of Interest}

The authors have no conflicts of interest to disclose.

\section{Data Availability}

The data that support the findings of this study are available from the corresponding author upon reasonable request.

\bibliography{aipsamp}

\end{document}